\documentclass[twocolumn,10pt]{IEEEtran}
\usepackage{bm,cite,float,amsmath,amssymb,amsthm}
\usepackage{subfigure}
\usepackage[noend]{algpseudocode}
\usepackage{indentfirst}
\usepackage[table]{xcolor}

\usepackage[linesnumbered,boxed,ruled,commentsnumbered]{algorithm2e}
\usepackage{algpseudocode}  
\usepackage{amsmath}  
\floatname{algorithm}{algorithm}
\renewcommand{\algorithmicrequire}{\textbf{Input:}}  % Use Input in the format of Algorithm  
\renewcommand{\algorithmicensure}{\textbf{Output:}} % Use Output in the format of Algorithm

\makeatletter
\def\BState{\State\hskip-\ALG@thistlm}
\makeatother

\makeatletter
\newcommand{\removelatexerror}{\let\@latex@error\@gobble}
\makeatother

\usepackage{amssymb}
\usepackage{amsmath}
\usepackage{graphicx}
\usepackage{cite}
\usepackage{balance}
\usepackage[utf8]{inputenc}

\IEEEoverridecommandlockouts

\usepackage{graphicx,epstopdf}
\usepackage{epsfig}	
\usepackage{amsfonts,balance}
\usepackage{bbm}

\usepackage[justification=centering]{caption}

\floatname{algorithm}{Algorithm}
\usepackage{lipsum}

\newtheorem{theorem}{Theorem}

\newtheorem{corollary}{Corollary}

\makeatletter
\def\ScaleIfNeeded{%
	\ifdim\Gin@nat@width>\linewidth \linewidth \else \Gin@nat@width
	\fi } \makeatother

\begin{document}
	
	\title{%Multiple Configured Grants Framework in Grant-Free NOMA for Massive URLLC
		%Configured-Grants Optimization in Grant-Free NOMA for Massive URLLC
		\textcolor{red}{Optimization of Grant-Free NOMA with Multiple Configured-Grants for mURLLC}
	}
	
	%Deep Reinforcement Learning-based Grant-Free Non-Orthogonal Access for  URLLC Service

	\author{ Yan Liu, ~\IEEEmembership{Member,~IEEE,}
		Yansha Deng,
		~\IEEEmembership{Member,~IEEE,}\\
		Maged Elkashlan,~\IEEEmembership{Senior Member,~IEEE,}
		Arumugam Nallanathan, ~\IEEEmembership{Fellow,~IEEE,}\\
		and George K. Karagiannidis, ~\IEEEmembership{Fellow,~IEEE}
		
		\vspace{-0.2cm}		 
		\thanks{Y. Liu is with the Key Laboratory of Ministry of Education in Broadband
			Wireless Communication and Sensor Network Technology, Nanjing University
			of Posts and Telecommunications, Nanjing 210003, China and with School of Electronic Engineering and Computer Science, Queen Mary University of London, London, UK
			(e-mail: yan.liu@qmul.ac.uk). }
		\thanks{M. Elkashlan, and A. Nallanathan are with School of Electronic Engineering and Computer Science, Queen Mary University of London, London, UK
			(e-mail:\{maged.elkashlan, a.nallanathan\}@qmul.ac.uk). }
		\thanks{ Y. Deng is with Department of Engineering, King's College London, London, UK (e-mail: yansha.deng@kcl.ac.uk).
			(Corresponding author: Yansha Deng)}
		\thanks{G. K. Karagiannidis is with Department of Electrical and Computer
			Engineering, Aristotle University of Thessaloniki, Thessaloniki 54 124, Greece
			(e-mail: geokarag@auth.gr).}
		\vspace*{+0.3cm}
	}

	\maketitle
	
	%\vspace*{-2cm}

	\begin{abstract}
		
		Massive Ultra-Reliable and Low-Latency Communications  (mURLLC), which integrates URLLC with massive access, is emerging as a new and important service class in the next generation (6G) for  time-sensitive traffics and has recently received tremendous research attention.
		However, realizing efficient, delay-bounded, and reliable communications for a massive number of user equipments (UEs) in mURLLC, is extremely challenging as it needs to simultaneously take into account the latency, reliability, and massive access requirements.
		To support these requirements, the third generation partnership project (3GPP) has introduced enhanced grant-free (GF) transmission in the uplink (UL), with multiple active configured-grants (CGs) for URLLC UEs. With multiple CGs (MCG) for UL, UE can choose any of these grants as soon as the data arrives.
		In addition, non-orthogonal multiple access (NOMA) has been proposed to synergize with GF transmission to mitigate the serious transmission delay and network congestion problems. 
		In this paper, we develop a novel learning framework for MCG-GF-NOMA systems with bursty traffic.
		We first design the  MCG-GF-NOMA  model by characterizing each CG using the parameters: the number of  contention-transmission units (CTUs), the starting slot of each CG within a subframe, and the number of repetitions of each CG. 
		Based on the model, the latency and reliability performances are characterized. 
		We then formulate the MCG-GF-NOMA resources configuration problem taking into account three constraints. 
		Finally, we propose a  Cooperative  Multi-Agent  based Double Deep  Q-Network  (CMA-DDQN) algorithm  to  balance the   allocations of the channel resources among  MCGs  so as  to   maximize the number of  successful  transmissions under the latency constraint. 
		Our results show that the MCG-GF-NOMA framework can simultaneously improve the low latency and high reliability performances in massive URLLC.

	\end{abstract}
	
	\begin{IEEEkeywords}
		Multiple configured-grants, massive URLLC, NOMA, deep reinforcement learning, resource configuration.
	\end{IEEEkeywords}
	
	\section{Introduction}
	% The Internet of Things (IoT) is an emerging and promising technology that tends to revolutionize the global world via heterogeneous smart devices through seamless connectivity \cite{Ericsson}\cite{7397856}. 
	% The third Generation Partnership Project (3GPP) standardization of cellular networks is trying to address the requirements of novel IoT use cases, in order to ensure that the technology standards evolve in addressing future market needs.
	% %It has become clear that the breadth of IoT use cases cannot be described with a simple set of cellular IoT requirements. 
	In the standardization of the Fifth  Generation  (5G)  New  Radio  (NR), three communication service categories were defined  to address the requirements of novel Internet of Things (IoT) use cases \cite{series2015imt}.
	%Two of them, massive Machine Type Communications (mMTC) and Ultra-Reliable and Low Latency Communications (URLLC), are specifically designed for future IoT applications\cite{3gpp2018study}.
	Among them, the Ultra-Reliable and Low-Latency Communications (URLLC) is one of the most challenging services with stringent low latency and high reliability requirements, i.e., in the Third Generation Partnership Project  (3GPP) standard \cite{3gpp2018study}, a general URLLC requirement is $1-10^{-5}$ target reliability within 1 ms user plane latency\footnote{User plane latency is defined as the one-way radio latency from the processing of the packet at the transmitter to when the packet has been received successfully and includes the transmission processing time, transmission time and reception processing time.}. 
	Considering the explosive increase in the number of IoT devices, it is essential to improve the access performance in networks for accommodating massive access with various requirements. 
	% The massive URLLC (mURLLC), which integrates the URLLC with massive access, is becoming a new and important service class in the next generation (6G) for the time-sensitive traffics and has received tremendous research attention\cite{9186090}.
	% However, realizing efficient, delay-bounded, and reliable communications for a massive number of IoT devices in mURLLC, is extremely challenging as it needs to take the latency, reliability, and massive access requirements into account at the same time.
	Integrating  URLLC with massive access, massive URLLC (mURLLC) wireless networks are able to realize efficient, delay-bounded, and reliable communications for a massive number of IoT devices\cite{9186090}.
	The mURLLC is becoming a new and important service class in the next generation (6G) for the time-sensitive traffics and has received tremendous research attention\cite{9398935}.
	However, addressing the need in mURLLC is fundamentally challenging as it needs to simultaneously guarantee the latency, reliability, and massive access requirements.
	%It is also anticipated that the device density may grow to hundred(s) of devices per cubic meter in the 6G white paper\cite{latva2020key}
	%To satisfy the requirements of emerging applications, such as intelligent transportation \cite{8246845}, industrial automation\cite{7497764}, and augmented/virtual reality (VR/AR)\cite{8663985} with high reliability and low latency, the third Generation Partnership Project (3GPP) has introduced URLLC as one of three service categories in the fifth generation (5G) New Radio (NR). 
	%The general requirements in URLLC design have been defined in 3GPP as: $1-10^{-5}$ reliability within 1ms user plane latency\footnote{User plane latency is defined as the one-way radio latency from the processing of the packet at the transmitter to when the packet has been received successfully and includes the transmission processing time, transmission time and reception processing time.} 
	%for 32 bytes (0.5ms for both downlink (DL)  and uplink (UL))\cite{3gpp2018study}.
	%Meanwhile, 6G wireless systems is advocating $1-10^{-9}$ reliability with 0.1 ms latency for supporting intelligent systems built upon
	%various perceptual modalities (or Internet of senses) and real-time human-machine interactions
	%\subsection{State-of-the-Art}
	
	To support these requirements, several new features such as configured-grant (CG) transmission with automatic repetitions\cite{3gpp38214}, user-equipment (UE) multiplexing\cite{8316582},  and multiple active CGs for URLLC UEs\cite{9305191} were standardized by the 3GPP.
	\subsubsection{Grant-Free NOMA}
	To reduce the latency in URLLC, the grant-free (GF) (a.k.a. configured-grant (CG)) transmission is proposed for 5G NR in 3GPP Release 15\cite{3gpp38214}  as an alternative for traditional grant-based (GB) (a.k.a. dynamic-grant (DG)) in Long Term Evolution (LTE).
	In NR GF transmission, the UE  is allowed to transmit data to the Base Station (BS) in an arrive-and-go manner without scheduling request (SR) and \textcolor{red}{uplink (UL)} resource grant (RG) to reduce latency. 
	To increase the reliability in URLLC, the  K-repetition GF transmission has been proposed by 3GPP, where a pre-defined number of  consecutive  replicas  of  the  same  packet  are  transmitted in the consecutive  time slots\cite{3gpp38214}. More details about K-repetition GF transmission can be found in \cite{9174916}.
	To mitigate the serious transmission delay and network congestion problems caused by collision events in contention-based GF transmission and enhance the uplink connectivity, non-orthogonal multiple access (NOMA) has been proposed to synergize with GF transmission \cite{8316582,9022993}, 
	where GF-NOMA allows multiple UEs to transmit over the same physical resource by employing user-specific signature patterns (e.g, codebook, pilot sequence, mapping pattern, demodulation reference signal, power, etc.)\cite{9097306}.
	% where
	% GF-NOMA allows multiple UEs to share the same time-frequency resource block (RB) via power domain or code domain multiplexing\cite{9097306} to 
	% since
	% a large number of access requests using the same time-frequency resources at the same
	% time can cause severe preamble collisions, resulting in serious transmission delay and network congestion \cite{9022993}.
	% GF-NOMA allows multiple UEs to share the same time-frequency resource block (RB) via power domain or code domain multiplexing\cite{9097306}.

	\subsubsection{Multiple Configured-Grants for Grant-Free NOMA}
\begin{figure}[htbp!]
		\centering
		\includegraphics[width=3.6in,height=2.4in]{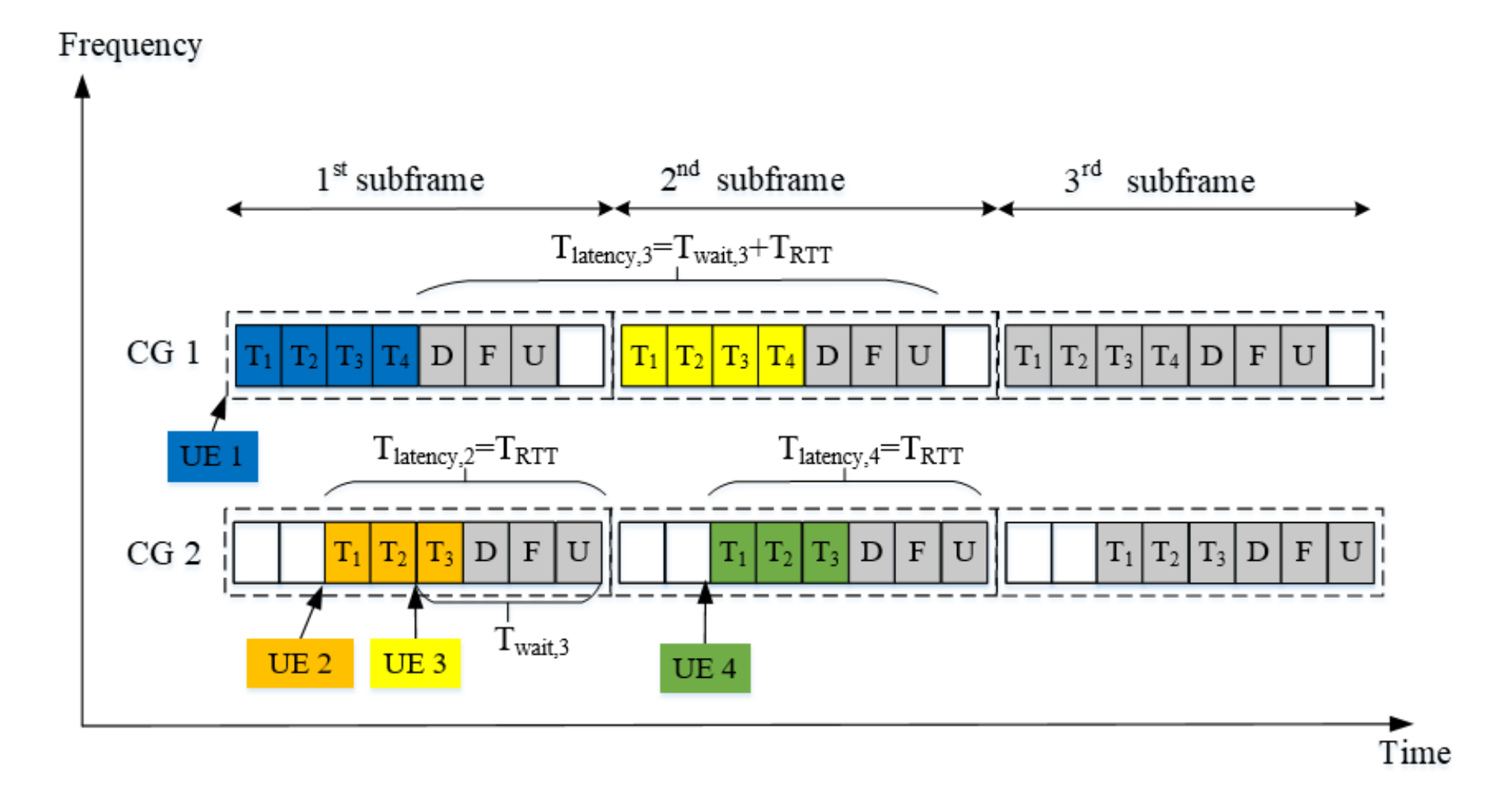}
		\caption{Multiple CGs (MCG) configurations for K-repetition GF transmission, T: packet  transmission, D: DL processing, F: ACK/NACK  feedback, and U: UL processing.
		}
		\label{fig:3}
	\end{figure}

	3GPP proposed multiple CGs (MCG) transmission in Release 16 \cite{9305191} to support different starting offsets of the resources with respect to UL packet arrival time  as shown in Fig.~\ref{fig:3}.
	On the one hand,  there is a chance of reducing the latency in cases where the data of an UE arrives (i.e., UE is active) after the starting slot offset of the CG 1 (UE 2, 3, and 4 in Fig.~\ref{fig:3}). 
	As illustrated in Fig.~\ref{fig:3}, UE 2 can transmit using the CG 2 without waiting for the CG period in the next subframe as in the single CG (SCG).
	% UEs do not have to wait for the CG period in the next subframe as in the single CG (SCG) to transmit the packet, which will bring additional latency.
	% For example, UE 2 needs wait for $T_{{\rm wait},2}$ TTIs and transmits at the beginning of next subframe.
	On the other hand, there is a chance of mitigating the collision events when multiple UEs are active and waiting for the CG period to transmit the packet.
	For example, UE 2 and UE 3  can transmit using different CG resources without collision as shown in Fig.~\ref{fig:3}.
	% \begin{figure}[htbp!]
		% 	\centering
		% 	\includegraphics[width=5.6in,height=3in]{single_CG2.pdf}
		% 	\caption{Single CG (SCG) configuration for K-repetition GF transmission, T: packet  transmission, D: DL processing, F: ACK/NACK  feedback, and U: UL processing.
			% 	}
		% 	\label{fig:2}
		% \end{figure} 
	% However, with this scheme, there are two problems.
	% On the one hand,  there is a chance of increasing the latency in cases where the UL data of an UE arrives (i.e., UE is active) after the starting slot offset of the CG (UE 2, 3, and 4 in Fig.~\ref{fig:2}). 
	% In such cases, UE has to wait for the CG period in the next subframe to transmit the packet which will add to the latency.
	% For example, UE 2 needs wait for $T_{{\rm wait},2}$ TTIs and transmits at the beginning of next subframe.
	% On the other hand, there is a chance of increasing the collision events where multiple UEs are active and waiting for the CG period to transmit the packet (UE 2 and 3 have collided using the same physical resource in Fig.~\ref{fig:2}). 
	% To mitigate these two problems, and as a part of enhancements to Release 15 UL URLLC specification, 3GPP proposed multiple CGs (MCG) transmission scheme in Release 16 to support different starting offsets of the resources with respect to UL packet arrival time \cite{9305191} as shown in Fig.~\ref{fig:3}. 
	Multiple CGs also support different resource sizes, repetitions, and periodicity, to suit different data requirements, respectively\cite{R1-1906151,9322288}. 
	
	%Multiple studies are done to evaluate the interference caused due to multiple UEs choosing the same CG [10]. Even if the collision is handled, there is no method which talks about how UE should choose a particular CG out of all CGs which makes sure that the latency and reliability criteria are met jointly
	
	%One configuration might be shared among several UEs.
	%One UE might have multiple CGs configurations and  
	%\subsection{Related Works}
	
	\subsection{Related Works}
	
	Scanning the open literature, to the best of our knowledge, most works focused on the analysis or optimization of single configured-grant GF-NOMA (SCG-GF-NOMA) transmissions.
	
	%Some recent academic works have foreseen the feasibility of applying  NOMA to single CG GF transmissions.
	In terms of analysis, a GF-NOMA strategy was proposed in \cite{7972955}, in which active devices transmitted data over a randomly selected available channel. In order to allow
	the receiver decode successfully, the transmitted data was encoded with rateless code.
	In \cite{8533378}, a new
	GF-NOMA analytical framework was proposed and the expressions for outage probability and throughput for GF-NOMA transmissions were derived, by treating collisions as interference through successive joint decoding or successive interference cancellation (SIC). In \cite{8662677}, a semi-GF scheme has been proposed, where the dedicated GB access was provided for one user while GF access was used by other users.

	In terms of optimization, several studies have applied deep reinforcement learning (DRL) to optimize the SCG-GF-NOMA networks.
	% Applying partial network observations and uniform resource access probabilities expropriate the conventional optimization approaches for SCG-GF-NOMA transmission, especially for long-term communications with time-varying channels. 
	DRL can obtain better resource allocation with near-optimal resource access probability distribution to improve the SCG-GF-NOMA transmission\cite{8986647}. 
	In \cite{8986647}, the authors designed users and sub-channel clusters in a region 
	to reduce collisions of the GF-NOMA
	system. The formulated long-term cluster throughput problem is solved via DRL algorithm for optimal sub-channel and power allocation.
	% The authors in [16] introduced deep RL
	% algorithm to improve the throughput of the grant-free NOMA
	% system. In order to reduce the computational complexity of
	% the model, the idea of subchannel and device clustering was
	% adopted.
	In \cite{9119119}, the authors
	introduced power-domain NOMA to further improve network throughput and defined a new reward that enabled only one acknowledgement bit returning to the device from the BS in each time slot.
	In \cite{9427159}, the authors proposed two
	distributed Q-learning aided uplink GF-NOMA schemes to maximize the number of
	accessible devices, where the bursty traffic of massive Machine  Type  Communications  (mMTC)  devices is carefully considered. 
	% In \cite{8625480}, DL was used to solve a variational optimization problem for SCG-GF-NOMA. 
	% The neural network model includes encoding, user activity, signature sequence generation, and decoding. 
	% The authors then extended their work 
	% to design a generalized/unified framework for NOMA using deep multi-task learning in \cite{8952876}. 
	% A DL-based active user detection scheme has been proposed for SCG-GF-NOMA in \cite{8968401} .
	% By feeding training data into the designed deep neural network, the proposed active user detection scheme learns the nonlinear mapping between received NOMA signal and indices of active devices.
	% Their results shown that the trained network can handle the whole
	% active user detection process, and achieve accurate detection
	% of the active users. 
	% These works assumed that each UE is pre-allocated with a unique sequence, and thus collisions are not an issue. However, this assumption does not hold in massive UEs settings in mURLLC, where the collision is the bottleneck of the GF-NOMA performance. 
	% Different from \cite{8986647,8625480,8952876,8968401}, we aim to develop a general learning framework to optimize  MCG-GF-NOMA systems for  massive URLLC taking into account the  MA signature collision, the UE detection as well as the data decoding procedures.
	
	Different from \cite{7972955,8533378,8662677,8986647,9119119,9427159}, we aim to first design a novel framework about multiple CGs GF-NOMA (MCG-GF-NOMA) networks and  optimize the long-term successfully served UEs under the latency constraint based on this framework for mURLLC service.

	\begin{figure*}[htbp!]
		\centering
		\includegraphics[width=6.5in,height=4.2in]{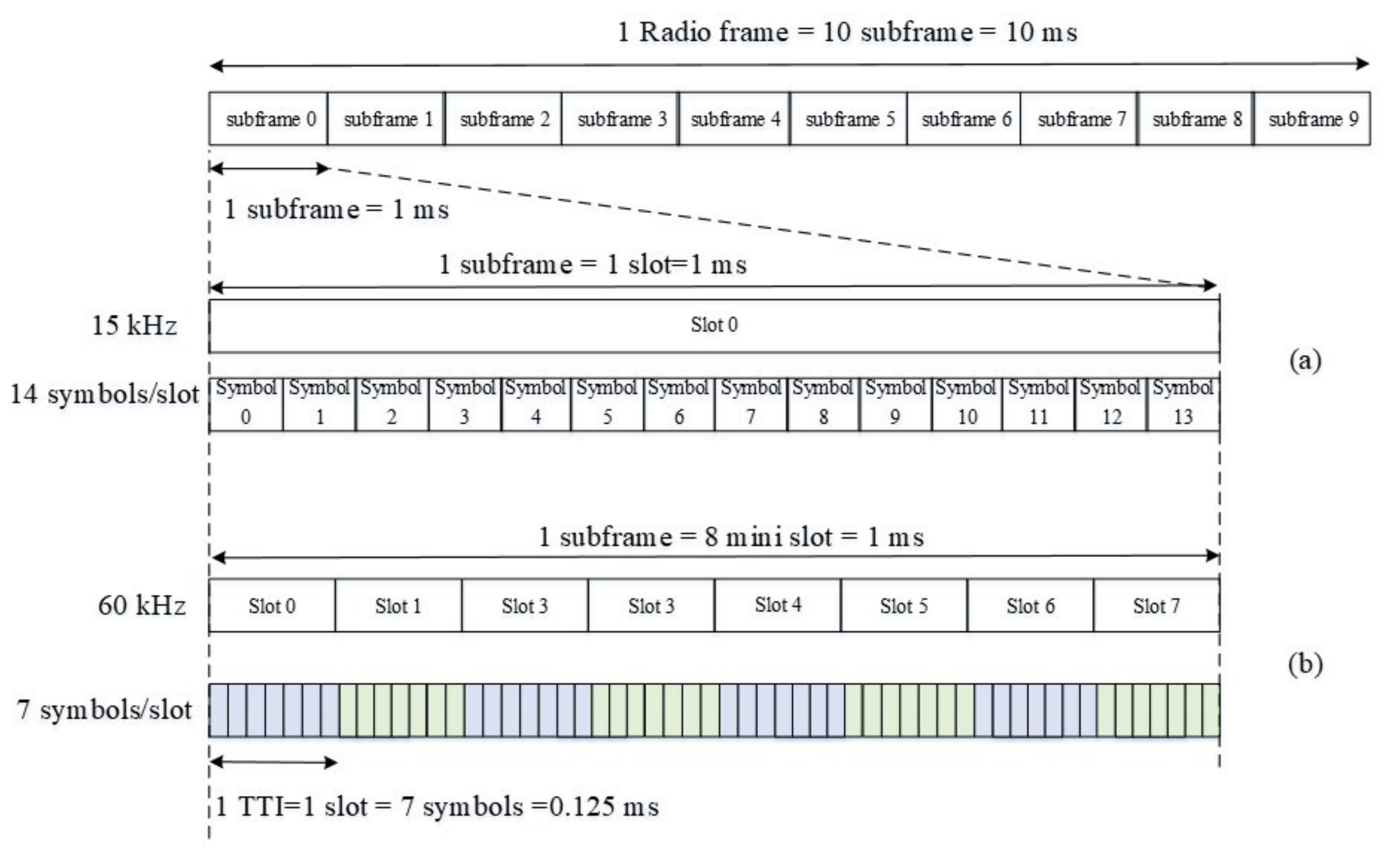}
		\caption{5G NR frame structure for numerology: (a) 15 kHz with 14 symbols/slot, (b) 60 kHz with 7 symbols/mini-slot.
		}
		\label{fig:1}
	\end{figure*} 
	
	\subsection{Motivations and Contributions}
	
	%It is worth noting that all of the above works have focused on the SCG-GF-NOMA systems. In addition, most of the existing  SCG-GF-NOMA systems  were  designed  mainly  to  maximize  the  thoughput of the mMTC.
	As mentioned before, research  on the  MCG-GF-NOMA networks  to support mURLLC  is  fundamental  and  essential,  which is an untreated and challenging problem.
	To cope with it, accurately  modeling, analyzing, and optimizing the MCG-GF-NOMA resource is fundamentally important, but the interplay between latency and reliability brings extra complexity.
	\textcolor{red}{In addition, in the GF-NOMA scheme, the data is transmitted along with the pilot randomly, which is unknown at the BS and can lead to new research problems. The blind detection of active UEs is needed due to that the set of active users is unknown to the BS, which also brings extra challenges.}
	\textcolor{red}{The MCG-GF-NOMA system optimization can hardly be solved via the traditional convex optimization method, due to the complex communication environment with the lack of tractable mathematical formulations, whereas  Reinforcement Learning (RL),  can be a potential alternative approach, due to that it solely relies on the self-learning of the environment interaction, without the need to derive explicit optimization solutions based on a complex mathematical model.}
	In this paper, we address the following fundamental questions:
	1) how to design the MCG-GF-NOMA network;
	2) how to quantify the URLLC reliability and latency performances in the MCG-GF-NOMA network;
	3) how to formulate the MCG-GF-NOMA resources configuration problem taking into account the reliability and latency;
	and 4) how to balance the allocations of channel resources among multiple CGs so as to provide maximum success transmissions in mURLLC scenario with bursty traffic.
	%To do so, we present a general learning framework for MCG-GF-NOMA networks in mURLLC.
	% \subsection{}
	The main contributions of this paper are as follows:
	
	\begin{itemize}
		\item We propose a novel MCG-GF-NOMA learning framework for attaining the long-term successfully served UEs under the latency constraint in mURLLC service, where 
		the latency and reliability performances are characterized and analyzed for each CG.
		In this framework,  we  practically  simulate  the  random traffics, the   resource   configuration, the collision detection, and the data decoding procedures.
		%We use this generated simulation environment to train the DRL agents.

		\item We design a  MCG-GF-NOMA  system, where we characterize each CG using the parameters including the number of  contention-transmission units (CTUs), the starting slot of each CG within a subframe, and the number of repetitions of each CG. We then formulate the MCG-GF-NOMA resource configuration problem taking into account three constraints: 1) 
		the  CTU resource constraint is set  to compare the MCG-GF-NOMA scheme with the SCG-GF-NOMA scheme; 2) the latency constraint is set to satisfy the latency requirement; and 3) the starting slot constraint is set to support various UL packet arrival times.
		% Thus, the  MCG-GF-NOMA configuration do not require the additional resources compared to the conventional SCG-GF-NOMA scheme.

		\item We propose a  Cooperative  Multi-Agent  learning  technique  based   Double Deep  Q-Network  (CMA-DDQN) algorithm  to  balance the   allocations of resources among  MCGs  so as  to   maximize the number of  successful  transmissions under the latency constraint, which breaks  down  the  selection  of  high-dimensional  parameters  into  multiple  parallel  sub-tasks with a number of DDQN agents cooperatively being trained to produce each parameter. 
		
		\item Our results show that the MCG-GF-NOMA learning framework can improve the low latency and high realibity performances in a massive URLLC scenario. First, the number of successfully served UEs in the MCG-GF-NOMA system is up to four times more than that in the SCG-GF-NOMA system, and the latency of successfully served UEs in the MCG-GF-NOMA system is circa half of that in the SCG-GF-NOMA system. Second, the MCG-GF-NOMA learning framework can also increase the CTU  resource utilization efficiency compared to the SCG-GF-NOMA system.
	\end{itemize}

	\subsection{Organization}
	The remainder of this paper is structured as follows.
	Section II illustrates the system model of MCG-GF-NOMA system.
	Section III describes the  problem analysis and formulation.
	Section IV elaborates on the proposed CMA-DDQN algorithm for solving the formulated problem. 
	The simulation results are illustrated in Section V. Finally, Section VI concludes
	the main concept, insights and results of this paper.

	\section{System Model}

	% \begin{figure}[htbp!]
		% 	\centering
		% 	\includegraphics[width=5.4in,height=2.8in]{one CG example.PNG}
		% 	\caption{NR PUSCH URLLC transmission using a CG with $N_{\rm RB}=1$, $N^{\rm start}_{\rm start}=2$, $N_{\rm repe}^{}=2$,  and $N_{\rm start}^{\rm sym}=7$
			% 	}
		% 	\label{fig:1}
		% \end{figure} 
	
	We consider a single-cell uplink wireless network with a coverage radius of $R$. Particularly,  a BS is located at the center of the cell, and a number of $N_{\rm UE}$ static UEs are randomly distributed around the BS in an area of the plane $\mathbb{R}^2$, where the UEs remain spatially static once deployed.
	The  BS is unaware of the status of these UEs, hence no uplink channel resource is scheduled to them in advance. 
	%In each IoT device uplink data is generated according to random inter-arrivalprocesses over the TTIs, which are Markovian and possibly time-varying.
	% The time is divided into TTIs, which refers to a mini-slot in this paper as shown in Fig.~\ref{fig:1}. The 5G NR introduces the concept of `mini-slots' and supports a scalable numerology allowing the sub-carrier spacing (SCS) to be expanded up to 240 kHz. 
	% Collectively, mini-slots and flexible numerology  allow shorter transmission slots to meet the stringent latency requirement. 
	To capture the effects of the physical radio,
	we consider the standard power-law path-loss model with the path-loss attenuation $r^{-\eta}$, where $r$ is the Euclidean distance between the UE and the BS and $\eta$ is the path-loss attenuation factor. 
	In addition, we consider a Rayleigh flat-fading environment, where the channel power
	gains $h$ are exponentially distributed (i.i.d.) random variables with unit mean.

	\subsection{5G NR Frame Structure and Numerologies}

	5G NR defines five numerologies based on subcarrier spacing (SCS) $\Delta f = {2^\mu } \times 15$ kHz, where $\mu=0,1,2,3,4$ is the numerology factor
	% 15 kHz, 30 kHz, 60 kHz, 120 kHz, and 240 kHz
	\cite{3gpp38211}, instead of a single value of 15 kHz in LTE.
	This feature reduces transmission time by decreasing the slot length as shown in Fig.~\ref{fig:1}. 
	As depicted in Fig.~\ref{fig:1}, the per
	frame duration in NR is still 10 ms, and the same as in LTE. One frame consists of 10 subframes and each with 1 ms duration.
	With the increased SCS, i.e., a large value of $\mu$, the slot duration reduces according to $1/2^{\mu}$ ms.
	To further reduce the latency by shortening transmission time interval (TTI), in 5G NR, a TTI can be a mini-slot of 2, 4, or 7 Orthogonal Frequency Division Multiplexing (OFDM) symbols instead of 14 OFDM symbols per TTI in LTE (see Fig.~\ref{fig:1}), and a transmission can start at the beginning of a mini-slot \cite{3gpp38211}.
	% Fig.~\ref{fig:1} shows the frame structurea and TTI duration for 15 kHz and 60 kHz SCS.
	Mini-slot durations  will depend on the SCS ($\mu$) and on the number of OFDM symbols included in a slot ($N_{\rm sym}$),  i.e., 
	
	\begin{align}\label{slot duration}
		{\rm TTI}=N_{\rm sym}/2^{\mu}/14 \ (\rm ms).
	\end{align}
	Thus, one NR subframe may have one (for $\mu=0$)
	or multiple slots depending on the value of the numerology factor $\mu$, i.e.,
	
	\begin{align}\label{slot number}
		N_{\rm slot}=1/{\rm TTI}=2^{\mu}\times14/N_{\rm sym}.
	\end{align}

	\subsection{Inter-Arrival Traffic}
	The small packets for each UE are generated according to random inter-arrival processes over the TTIs, which are Markovian as defined in \cite{name2015,3gpp37868} and unknown to BS.
	We consider a bursty traffic process, which occurs when a large number of UEs attempt to access the same network simultaneously during a short period of time\cite{8985528}. 
	This is especially observable when the number of UEs could be huge. 3GPP recommends applying a Beta distribution based arrival process to model the arrival intensity during bursty traffic arrivals in \cite{3gpp37868}.
	Considering the nature of slotted-Aloha, the newly activated devices can only execute transmission at the beginning of the closest CG. 
	This means that the UEs transmitting in a ${\rm CG}_i$ period
	come from those who received a packet within the interval between the last period ($\tau^{i-1}$,$\tau^{i}$).
	The traffic instantaneous rate in packets in a period is described by a function $p(\tau)$, so that the packets arrival rate in the $i$th CG period  is given by
	\begin{align}\label{rate}
		{A ^i} = \int_{{\tau _{i - 1}}}^{{\tau _i}} {p(\tau )} d\tau.
	\end{align}
	Each UE would be activated at any time $\tau$, according to a time limited Beta probability density function (PDF) as \cite[Section 6.1.1]{3gpp37868}
	\begin{align}\label{beta}
		p(\tau) = \frac{{{\tau ^{\alpha  - 1}}{{(T - \tau )}^{\beta  - 1}}}}{{{T^{\alpha  + {\beta}  - 1}} \rm {Beta}(\alpha ,\beta )}},
	\end{align}
	where  $T$ is the total time duration of the bursty traffic and ${\rm Beta} (\alpha , \beta)=\int_0^1 {{\tau ^{\alpha  - 1}}} {(1 - \tau )^{\beta  - 1}}d\tau$ is the Beta function with the constant
	parameters $\alpha$ and $\beta$\cite{gupta2004handbook}.

	%\textcolor{red}{related the time duration}

	\subsection{Grant-Free NOMA Model} 
	
\begin{figure}[htbp!]
		\centering
		\includegraphics[width=3.7in,height=1.4in]{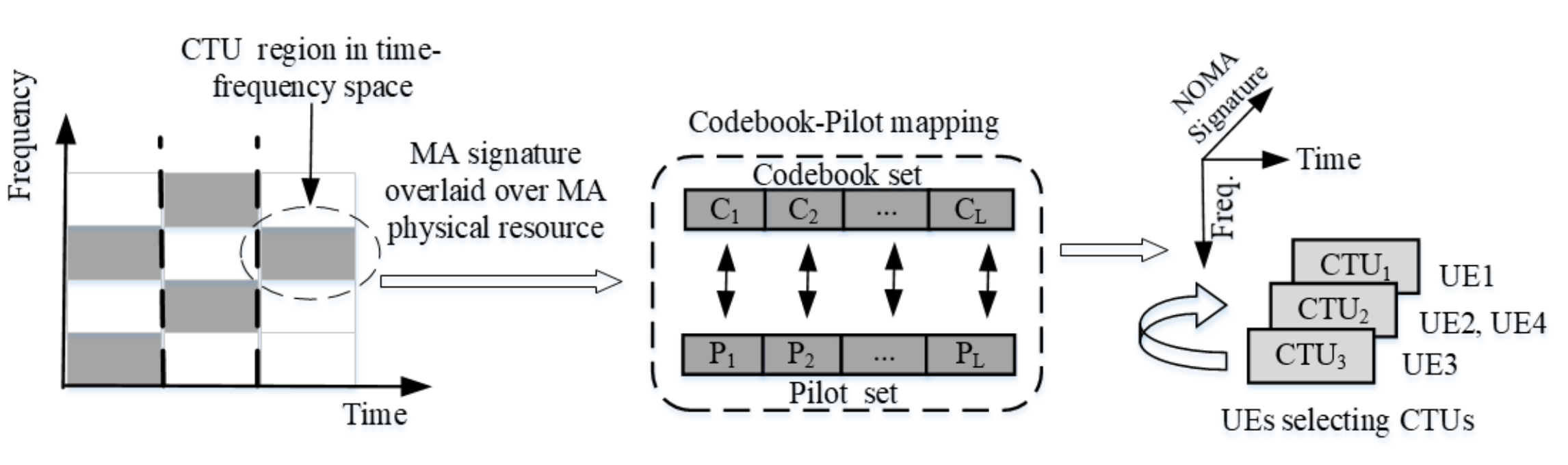}
		\caption{An illustration of CTU in a time-frequency space.
		}
		\label{fig:4}
	\end{figure}

	We focus on the UEs that are connected to the network in a GF manner. In order to deal with the resource constraint problem caused by orthogonal resource allocation, NOMA is introduced to increase the number of accessible devices in this paper.
	In the GF-NOMA, the smallest transmission unit that a UE can compete for is called a contention transmission
	unit (CTU). 
	A CTU may comprise of a MA physical resource and a MA signature \cite{ye2018uplink,9097306,7063547}.
	The MA physical resources  represent a set of time-frequency resource blocks (RBs) and the MA signatures represent a set of pilot sequences for channel estimation and/or UE activity detection, and a set of codebooks for robust data transmission and interference whitening, etc.
	Without loss of generality, we consider that there are $L$ different pilot sequences defined over one time-frequency RB as shown in Fig.~\ref{fig:4}.
	Each pilot sequence $l$ is made unique to a specific codebook and acts as the UE's signature\footnote{A one-to-one mapping or a many-to-one mapping between the pilot sequences  and codebooks can be predefined.
		Since it  has been verified in \cite{7887683} that the performance loss due to codebook  collision is negligible for a real system, we focus on the pilot sequence collision and consider the one-to-one mapping as \cite{8533378,9031550}.}\cite{8533378,8316582}.
	There are obviously $N_{\rm CTU}=F\times L$ unique CTUs over $F$ time-frequency RBs configured  by the BS in each CG configuration period.
	Each UE randomly choose one CTU from the pool to transmit in this
	period.
	Unlike orthogonal resource allocation (i.e., each time-frequency resource can only be used by one UE), NOMA allows multiple UEs with different codebooks and pilot sequences to transmit over the same time-frequency resource, thus increasing the number of accessible UEs without expanding physical resources. 
	However, a collision will occur when more than one UE selects the same codebook and pilot sequence (i.e. the same CTU).

	\begin{figure*}[htbp!]
		\centering
		\includegraphics[width=6.2in,height=4.0in]{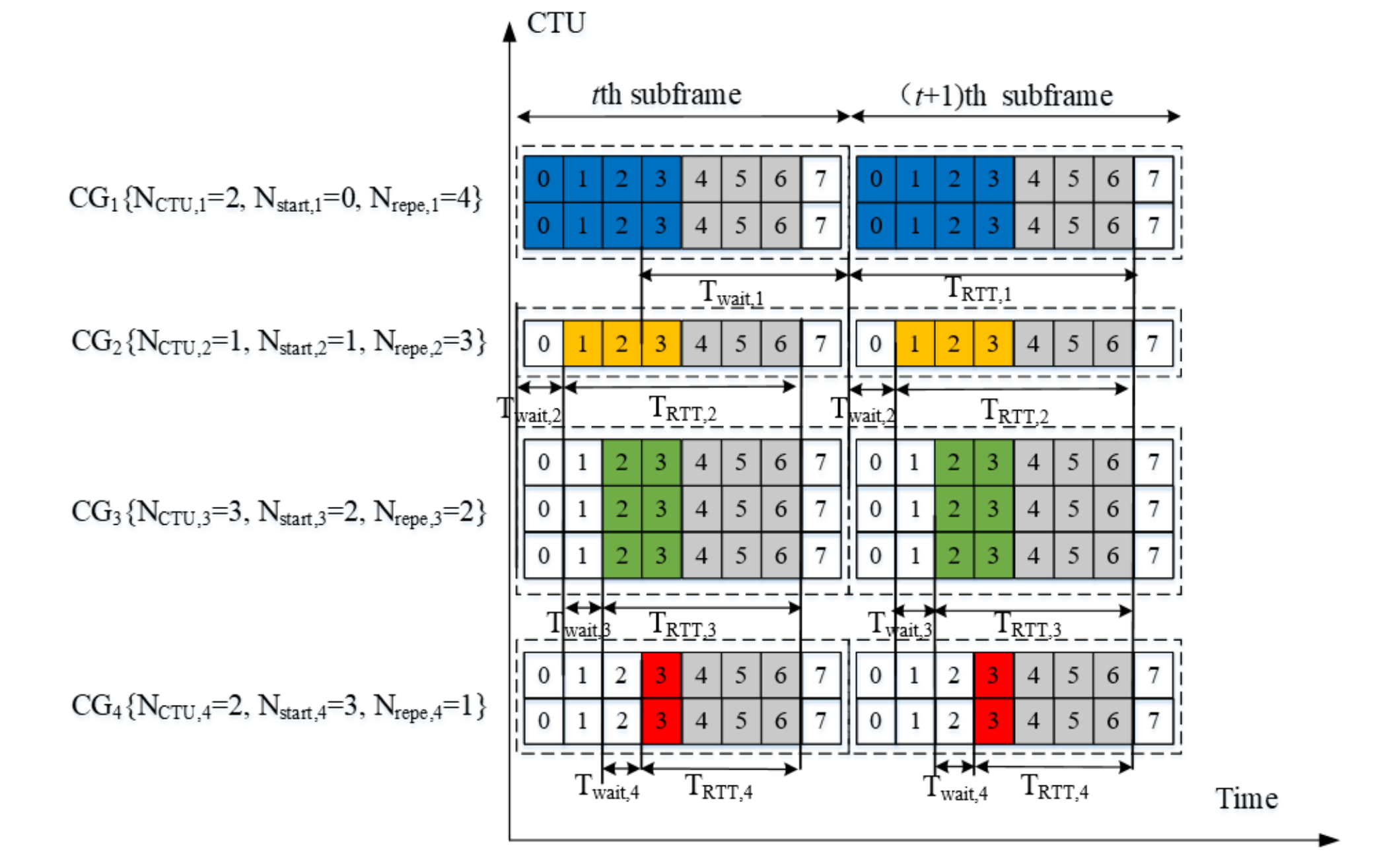}
		\caption{Multiple CGs (MCG) configurations with four CGs.
		}
		\label{fig:5}
	\end{figure*} 
	
	\subsection{Multiple Configured-Grants Grant-Free NOMA (MCG-GF-NOMA) Design}  
	
	We consider the MCG-GF-NOMA system as shown in Fig.~\ref{fig:5}.
	The BS configures $N_{\rm CG}$ UL CGs for massive URLLC transmissions at each subframe.
	The UE chooses the configuration with the earliest starting point to transmit data.
	% \begin{figure}[htbp!]
		% 	\centering
		% 	\includegraphics[width=6.2in,height=4.5in]{multiple_CG_B.PNG}
		% 	\caption{Multiple CGs configuration scheme B with five multiple CGs.
			% 	}
		% 	\label{fig:6}
		% \end{figure} 
	Each CG is consist of different resources in the CTU domains, and is associated with the following transmission parameters:
	\begin{itemize}
		\item Number of CTUs ($N_{\rm CTU}$)
		\item Starting slot  within a subframe ($N_{\rm start}$)
		\item Number of repetitions ($N_{\rm repe}$)
		%\item Numerology factor ($\mu$)
		\item Number of slots in a subframe ($N_{\rm slot}$)
		% \item Configuration period in a subframe ($N_{\rm start}^{\rm sym}$)
		%\item Number of OFDM symbols in a slot ($N_{\rm sym}$)
	\end{itemize}
	% Without loss of generality, we focus on the mini-slots of $N_{\rm sym}=7$ OFDM symbols for transmissions using 60 kHz ($\mu=2$) SCS, which is in line with the main guidelines for 3GPP NR performance evaluations presented in \cite{2Tel2018}.
	% Thus, the TTI duration is TTI $=N_{\rm sym}/2^{\mu}/14=0.125$ ms and the number of slots in a subframe is $N_{\rm slot}=1/{\rm TTI}=8$ TTIs.
	% The configuration of other numerology factors can be easily extended.
	Without loss of generality, we consider that all the subframe has the same number of slots all the time, i.e., the $N_{\rm slot}$ is the same for each CG and each subframe.
	Thus, for ease of presentation, we represent each ${{\rm CG}_i}$ in the $t$th subframe by ${\rm CG}^t_i\{N^t_{{\rm CTU},i}, N^t_{{\rm start},i}, N^t_{{\rm repe},i}\}$.
	As illustrated in Fig.~\ref{fig:5}, ${\rm CG}^t_1\{2, 0, 4\}$, ${\rm CG}^t_2\{1, 1, 3\}$, ${\rm CG}^t_3\{3, 2, 2\}$, and ${\rm CG}^t_4\{2, 3, 1\}$ are four CGs in the $t$th subframe.

	The main variables are summarized in
	Table~\ref{table:1}.

	\begin{table*}[htbp!]
		\centering
		\caption{Notation Table}
		{\renewcommand{\arraystretch}{1.2}
			%\renewcommand{\tabcolsep}{0.15cm}
			%	\begin{tabular}{|c|c|c|c|}
				\rowcolors{0}{gray!25}{white}
				\begin{tabular}{|p{1.5cm}|p{5.5cm}|p{1.5cm}|p{5.5cm}|}
					\hline
					
					$\textbf{Symbol}$ & \textbf{Meaning} & $\textbf{Symbol}$ & \textbf{Meaning}  \\ \hline
					
					$N_{\rm UE}$ & \textcolor{red}{The number of static UEs} & 
					$R$ & The coverage radius of the cell \\ \hline
					$r$ & The distance between an UE and the BS &  
					$h$ & The Rayleigh fading channel power gain  \\ 
					\hline
					$\eta$ & The path-loss exponent & 
					$\mu$&  The numerology factor  \\ \hline 
					$N_{\rm sym}$ & The number of OFDM symbols included in a slot & 
					$N_{\rm slot}$&  The number of slots within a subframe  \\ \hline
					$A$ & The packets arrival rate &
					$p$&  The Beta probability density   \\ \hline
					$T$ & The duration of the bursty traffic &
					$L$&  The number of pilot sequences over one RB  \\ \hline
					$F$ & The number of time-frequency RBs & 
					$N_{\rm CG}$ & The number of CGs configured at each subframe   \\ \hline
					$N_{\rm CTU}$ &  The number of CTUs &  
					
					${\cal N}_{\rm CTU}$ & The set of the number of CTUs    \\ \hline
					$N_{\rm start}$ & The starting slot  within a subframe & 
					${\cal N}_{\rm start}$ & The set of the starting slot    \\ \hline
					$N_{\rm repe}$ & The number of repetitions &  
					$t$ & The $t$th subframe   \\ \hline
					$T_{{\rm wait},i}^t$ & The waiting time of the UE using ${{\rm CG}_i}$ at the $t$th subframe &  
					$T_{{\rm RRT},i}^t$ & The RTT time of the UE using ${{\rm CG}_i}$ at the $t$th subframe \\ \hline
					$T_{{\rm laten},i}^t$ & The latency of the UE using ${{\rm CG}_i}$ at the $t$th subframe &  
					$T_{{\rm aver}}^t$ & The average latency of the successfully served UEs at  each subframe  \\ \hline
					$N_{{\rm suc},i}^t$ & The successfully served UEs using  ${{\rm CG}_i}$ at the $t$th subframe &  
					${\cal N}_{{\rm IC},i}^t$ & The set of idle CTUs for ${\rm CG}_i$ at the $t$th subframe  \\ \hline
					${\cal N}_{{\rm SC},i}^t$ & The set of singleton CTUs for ${\rm CG}_i$ at the $t$th subframe &  
					${\cal N}_{{\rm CC},i}^t$ & The set of collision CTUs for ${\rm CG}_i$ at the $t$th subframe  \\ \hline
					${\cal N}_{f,{\rm SU},i}^t$ & The  set  of  UEs  choosing  the  singleton  CTUs  for   ${\rm CG}_i$ on the $f$th RB &   ${N}_{f,{\rm CU},i}^t$ & The  number  of  UEs  choosing  the  collision  CTUs  for   ${\rm CG}_i$ on the $f$th RB  \\ \hline
					$P$ & The  transmission power   & 
					$\sigma^2$ & The  noise power  \\ \hline
					$\gamma_{th}$ & The received SINR threshold & $CG_i$ & The $i$th CG in a subframe\\ \hline
					$N_{{\rm CTU},{\rm SCG}}$ &  The configured CTU numbers for the SCG-GF-NOMA system & $T_{\rm RTT}$ & The length of one round trip time\\ \hline
					$T_{\rm wait}$ &  The length of waiting time & $T_{\rm laten}$ & The latency of the successfully served UE\\ \hline
					$T_{\rm aver}$ & The average latency of the successfully served UEs in each subframe & $P_{s,f,i}$ & The received power of  the $s$th UE in the $n$th repetition of the CG $i$ on the $f$th RB\\ \hline
				\end{tabular}
			}
			\label{table:1}
		\end{table*}

			\section{Problem Analysis and Formulation}
			
			%\subsection{GF-NOMA Latency and Reliability}
			
			In a given subframe $t$, the BS \textcolor{red}{preconfigured} $N_{\rm  CG}$ CGs for UEs to transmit their packets.
			The BS sends radio resource control (RRC) (for both type 1 and type 2 CG transmission) or downlink control information (DCI) (only for type 2 CG transmission) to activate or release the CG configurations\cite{3gpp38824}.
			As soon as the URLLC data arrives, a UE can choose the ${\rm CG}^t_i$ with the earliest starting point (i.e., the smallest $N^t_{{\rm start},i}$) to transmit data.
			Suppose that the UE choose the ${\rm CG}^t_i\{N^t_{{\rm CTU},i}, N^t_{{\rm start},i}, N^t_{{\rm repe},i}\}$, then the UE randomly choose a CTU from $N^t_{{\rm CTU},i}$ available CTUs and start transmit at slot $N^t_{{\rm start},i}$ for $N^t_{{\rm repe},i}$ repetitions.
			The BS decodes (D) each repetition independently and the transmission is successful when at least one repetition succeeds.
			After processing all the received $N^t_{{\rm repe},i}$ repetitions, the BS transmits the ACK/NACK feedback (F) to the UE.
			% As soon as the URLLC data arrives, a UE can choose the ${\rm CG}^t_i\{N^t_{{\rm repe},i}, N^t_{{\rm CTU},i}, N^t_{{\rm start},i}\}$ $(i=1,2,3,4)$ with the earliest starting point (i.e., the smallest $N^t_{{\rm start},i}$) to transmit data.
			Considering the small packets of URLLC traffic, we set the packet transmission  time as one TTI. The BS feedback time   and the BS (UE) processing time are also assumed to be one TTI following  our previous work \cite{9174916}.
			The latency analysis and the reliability analysis for the MCG-GF-NOMA are described in the following.

			%\subsection{Multiple CGs Configuration Scheme A}
			
			\subsection{MCG-GF-NOMA Latency Analysis}
			
			In order to meet the low latency requirement for mURLLC, we consider that the active UE can only transmit for one round trip time (RTT).
			The RTT is the length time it takes for a data packet to be sent to a destination plus the time it takes for an acknowledgment of that packet to be received back at the origin.
			According to Fig.~\ref{fig:5}, the incurred latency of the UE using the $CG_i^t$ at the $t$th subframe includes two parts: the waiting time $T_{{\rm wait},i}^t$ and the RTT $T_{{\rm RTT},i}^t$.
			We obtain the RTT of the UE using $CG_i^t$ at the $t$th subframe as
			\begin{align}\label{RTT}
				{T_{{\rm RRT},i}^t} =N^t_{{\rm repe},i} + 3.
			\end{align}
			It should be noted  that the UEs transmitting in  ${\rm CG}_i^t$ come from those who received a packet after the start point of the ${\rm CG}_{i-1}^t$.
			Thus, the waiting time is the length time from the start point of the ${\rm CG}_{i-1}^t$ to the start point of the ${\rm CG}_{i}^t$.
			We derive the waiting time as
			
			\begin{align}\label{wait}
				{T_{{\rm wait},i}^t} =\tau ^i-\tau ^{i-1},
			\end{align}
			where
			\begin{align}\label{packet interval}
				%\begin{array}{l}
				&({\tau ^{i-1}},{\tau ^i})=\\ \nonumber
				&\left\{ \begin{array}{l}
					( N_{\rm slot} \times (t - 1) + N^t_{{\rm start},{i-1}}, N_{\rm slot} \times (t - 1) + N^t_{{\rm start},i}),\\ \nonumber
					(i > 1), \\ \nonumber
					% ({\tau ^2},{\tau ^3}) = ( 8 \times (t - 1) + N^t_{{\rm start},2},8 \times (t - 1) + N^t_{{\rm start},3}), \\
					% ({\tau ^3},{\tau ^4}) = ( 8 \times (t - 1) + N^t_{{\rm start},3},8 \times (t - 1) + N^t_{{\rm start},4}), \\
					( 0,0), \\\nonumber
					(i=1,t = 1),\\\nonumber
					( N_{\rm slot} \times (t - 2) + N^t_{{\rm start},N_{\rm CG}^{t-1}}, N_{\rm slot} \times (t - 2) + N_{\rm slot}), \\\nonumber (i=1,t > 1).
				\end{array} \right.
				%\end{array}
			\end{align}
					According to \eqref{RTT}, \eqref{wait}, and \eqref{packet interval}, we obtain the latency for ${\rm CG}^t_i$ as
					\textcolor{red}{
						\begin{align}\label{latency}
							&{T_{{\rm laten},i}^t} = {T_{{\rm wait},i}^t} + {T_{{\rm RRT},i}^t}\\ \nonumber
							&=\left\{\begin{array}{l}
								N_{{\rm start},i}^t - N_{{\rm start},{i-1}}^t + N^t_{{\rm repe},i} + 3, (i>1),\\
								N^t_{{\rm repe},i} + 3, (i=1,t=1),\\
								% T_{{\rm latency},3}^t = N_{{\rm start},3}^t - N_{{\rm start},2}^t + N^t_{{\rm repe},3} + 3,\\
								% T_{{\rm latency},4}^t = N_{{\rm start},4}^t - N_{{\rm start},3}^t + N^t_{{\rm repe},4} + 3,\\
								N_{\rm slot} - N_{{\rm start},N_{\rm CG}^{t-1}}^{t} + N^t_{{\rm repe},i} + 3, (i=1,t>1).
							\end{array}  \right.
					\end{align}}
					%Substituting   \eqref{packet interval}  into   \eqref{latency}, we have 
					% \begin{align}\label{latency_i}
						% \left\{\begin{array}{l}
							% T_{{\rm latency},2}^t = N_{{\rm start},2}^t - N_{{\rm start},1}^t + N^t_{{\rm repe},2} + 3,\\
							% T_{{\rm latency},3}^t = N_{{\rm start},3}^t - N_{{\rm start},2}^t + N^t_{{\rm repe},3} + 3,\\
							% T_{{\rm latency},4}^t = N_{{\rm start},4}^t - N_{{\rm start},3}^t + N^t_{{\rm repe},4} + 3,\\
							% T_{{\rm latency},1}^t = 8 - N_{{\rm start},4}^t + N^t_{{\rm repe},1} + 3.
							% \end{array}  \right.
						% \end{align}

					In order to compare the latency performance, we calculate the average latency of the successfully served UEs in each subframe as
					
					\begin{align}\label{average_latency}
						T_{\rm aver}^t = \frac{{\sum\limits_i^{{N_{\rm CG}}} {T_{{\rm laten},i}^t \times N_{{\rm suc},i}^t} }}{{\sum\limits_i^{{N_{\rm CG}}} {N_{{\rm suc},i}^t} }},  
					\end{align}
					where $N_{{\rm suc},i}^t$ is the successfully served UEs using the ${\rm CG}_i$ at the $t$th subframe and is obtained in the next subsection about reliability analysis.
					
					\subsection{MCG-GF-NOMA Reliability Analysis}

					During each RTT, if the GF-NOMA procedure fails, the UE fails to be served  and its packets will be dropped.
					The GF-NOMA fails if: ($i$) a CTU collision occurs when two or more UEs choose the same CTU  (i.e., UE detection fails); or ($ii$) the SIC decoding fails (i.e., data decoding fails).
					
					\subsubsection{CTU dectection}
					
					%or ($iii$) the transmission latency $T_{\rm late}>T_{cons}$.
					At each RTT, each active UE transmits its packets to the BS by randomly choosing a CTU from the earliest ${\rm CG}_i$.
					The BS  can  detect the UEs that have chosen different CTUs.
					However, if multiple   UEs choose the same CTU, the BS cannot differentiate  these UEs and therefore cannot decode the data.
					We categorize the CTUs from each ${\rm CG}_i$ into three types\cite{8533378}: 
					\begin{itemize}
						\item \textit{idle} CTU: a CTU
						which has not been chosen by any UE;
						\item  \textit{singleton} CTU : a CTU chosen by only one UE;
						\item \textit{collision} CTU : a CTU chosen by two or more UEs.
					\end{itemize}
					After collision detection at the $t$th subframe for the ${\rm CG}_i$, the BS observes  the set of singleton CTUs  ${\cal N}_{{\rm SC},i}^t$, the set of idle CTUs  ${\cal N}_{{\rm IC},i}^t$, and the  set of collision CTUs   ${\cal N}_{{\rm CC},i}^t$ for each ${\rm CG}_i$.

					\subsubsection{SIC decoding}
					
					After detecting the UEs that have chosen the singleton CTUs, the BS performs the SIC technique to decode the data of these UEs.
					Based on the NOMA principles, at each iterative stage of SIC, the BS first decodes the UE with the strongest received power and then subtracted the successfully decoded signal  from the received signal (we assume perfect SIC  the same as \cite{8533378}).
					That is to say, the decoding order at the BS is in sequence to the received power.
					It worth noting that during the decoding, the UEs that transmit on different RBs do not interfere with each other due to the orthogonality, and  only UEs that transmit on the same RB cause interference.
					Thus, in order to characterize the UEs transmitting with ${\rm CG}_i$ on the $f$th RB, we represent the ${\cal N}_{f,{\rm SU},i}^t$ as the set of UEs that have chosen the singleton CTUs for the ${\rm CG}_i$ on the $f$th RB, the ${ N}_{f,{\rm SU},i}^t=\left| {\cal N}_{f,{\rm SU},i}^t \right|$ as the number of UEs that have chosen the singleton CTUs for the ${\rm CG}_i$ on the $f$th RB ($\left|  \cdot  \right|$ denotes
					the number of elements in any vector $\cdot$), and ${ N}_{f,{\rm CU},i}^t$ as the number of UEs that have chosen the collision CTUs using the ${\rm CG}_i$ on the $f$th RB.
					We define the received power of the $s$th UE in the $n$th repetition of the ${\rm CG}_i$ on the $f$th RB as
					\begin{align}\label{receive}
						P_{s,f,i}^t={{P}{h_{s,f,i}^t}{r_{s}}^{ - \eta }},
					\end{align}
					where $P$ is the  transmission power, $r$ is the Euclidean distance between the UE and
					the BS, $\eta$ is the path-loss attenuation factor, $h$ is  the Rayleigh fading channel power gain from the UE to the BS.

					Suppose that the received power obeys $P_{1,f,i}^t\ge P_{2,f,i}^t\ge...\ge P_{{ N}_{f,{\rm SU},i}^t}^t$, the decoding order should be from the $1$st UE to the ${ N}_{f,{\rm SU},i}$th UE.
					In each iterative stage of SIC decoding, the CTU with the strongest received power is decoded by treating the  received powers of other CTUs over the same RB  as the interference.
					Thus, at the $t$th subframe, in the $n$th repetition of the ${\rm CG}_i$ on the $f$th RB,  the signal-to-interference-plus-noise ratio (SINR) of the  $s$th stage of SIC decoding of the  $s$th UE is derived as
					\begin{align}\label{SIC}
						{\rm SINR}^{t}_{s,f,i}= \frac{{P_{s,f,i}^t}}{{\sum\limits_{m=s+1}^{ {  N}_{f,SU,i}^t} P_{m,f,i}^t + \sum\limits_{n' =1}^{{N}_{f,CU,i}^t} P_{n',f,i}^t + {\sigma ^2}}},
					\end{align}
					where $\sigma^2$ is the noise power.
					%In each iterative stage of SIC decoding, the CTU with the strongest received power is decoded by treating the  received powers of other CTUs over the same RB  as the interference.
					
					Each iterative stage of SIC decoding is successful when the SINR in that stage is larger than the SINR threshold, i.e., ${\rm SINR}^{t}_{s,f,i}\ge {\gamma _{th}}$.
					% It worth noting that during the decoding, the UEs that transmit on different RBs do not interfere with each other due to the orthogonality, and  only UEs that transmit on the same RB cause interference.
					% If the received signal is decoded successfully, the decoded signal is subtracted from the received signal\footnote{We assume perfect SIC  the same as \cite{8533378}, with no error propagation between iterations.}. 
					%In the second stage, the second strongest received power is decoded by regarding the remaining received powers as interference.
					% In \eqref{SIC}, $P$ is the  transmission power, $r$ is the Euclidean distance between the UE and
					% the BS, $\eta$ is the path-loss attenuation factor, $h$ is  the Rayleigh fading channel power gain from the UE to the BS, $\sigma^2$ is the noise power,
					% ${ N}_{f,{\rm SU},i}^t=\left\| {\cal N}_{f,{\rm SU},i}^t \right\|$ is the number of UEs that have chosen the singleton CTUs for ${\rm CG}_i$ on the $f$th RB, and${ N}_{f,{\rm CU},i}^t=\left\| {\cal N}_{f,{\rm CU},i}^t \right\|$ is the number of the UEs that have chosen the collision CTUs for ${\rm CG}_i$ on the $f$th RB.
					The SIC procedure stops when one iterative stage of the SIC fails or when there are no more signals to decode.
					The SIC decoding procedure for each ${\rm CG}_i$ is described in the following.
					\begin{itemize}
						\item Step 1: Start the $n$th repetition with the initial  $n=1$, ${\cal N}_{f,{\rm SU},i}^t$, ${N}_{f,{\rm SU},i}^t$ and ${ N}_{f,{\rm CU},i}^t$;
						
						\item Step 2: Decode the $s$th UE  with the initial  $s=1$ using \eqref{SIC};
						% Initialize the latency constraint $\cal T$=1 TTI, and $
						% {\cal P}_{F}^{\rm Reac}[T_{\rm latency}\leqslant{1}]=0$;
						\item Step 3: If the $s$th UE is successfully decoded, put the decoded UE in set ${\cal N}_{f,{\rm suc},i}^t(n)$ and go to Step 4, otherwise go to Step 5;
						\item Step 4: If $s \le { N}_{f,{\rm SU},i}^t$, do $s=s+1$, go to Step 2, otherwise go to Step 5;
						\item Step 5: SIC for the $n$th repetition stops;
						\item Step 6: If $n \le N_{{\rm repe},i}$, do $n=n+1$, go to Step 1, otherwise go to the end.
						% \item Step 5: Repeat the step 3 to 4 until $m=M$ and calculate the latent access failure probability $
						% {\cal P}_{\rm out}^{\rm Reac}[T_{\rm latency}\leqslant{\cal T}]$ using (\ref{krep_out}).
					\end{itemize}
					Finally, the set of successfully served UEs using the ${\rm CG}_i$ on the $f$th RB at the $t$th subframe is derived as
					\begin{align}\label{decoded UEs_i,f}
						{\cal N}_{f,{\rm suc},i}^t = \bigcup\limits_{n = 1}^{{N_{{\rm repe},i}}} {( {{\cal N}_{f,{\rm suc},i}^t(n)} )}, 
					\end{align}
					the set of the successfully served UEs using the ${\rm CG}_i$ at the $t$th subframe is obtained as 
					\begin{align}\label{decoded UEs_i}
						{\cal N}_{{\rm suc},i}^t = \bigcup\limits_{f = 1}^{{F^t}} {( {{\cal N}_{f,{\rm suc},i}^t} )},    
					\end{align}
					and the set of the successfully served UEs at the $t$th subframe is obtained as 
					\begin{align}\label{decoded UEs}
						{\cal N}_{{\rm suc}}^t = \bigcup\limits_{i = 1}^{{N_{\rm CG}}} {( {{\cal N}_{{\rm suc},i}^t} )}.    
					\end{align}
					Then, ${N}_{{\rm suc}}^t=\left| {{\cal N}_{{\rm suc}}^t } \right|$ is the number of successfully served UEs.

					\subsection{Problem Formulation}

					% Once actived in a given subframe $t$, a UE executes the MCG-GF-NOMA  procedure, where the UE randomly chooses one of the preconfigured $C^t$ CTUs to transmit its packets for $K_{\rm{Krep}}^t$ times or  $k_{\rm{Proa}}^t\le K_{\rm{Proa}}^t$ times under the K-repetition scheme  and the Proactive scheme, respectively. 
					% During this RTT, the GF-NOMA fails if: ($i$) a CTU collision occurs when two or more UEs choose the same CTU  (i.e., UE detection fails); or ($ii$) the SIC decoding fails (i.e., data decoding fails).
					% %or ($iii$) the transmission latency $T_{\rm late}>T_{cons}$.
					% Once failed, UEs decides whether to retransmit in the following RTT or not based on the transmission latency check. When $T_{\rm late}>T_{\rm cons}$, the UE fails to be served  and its packets will be dropped.
					% It is obvious that 1) increasing the repetition values  $K^t$ could improve the GF-NOMA success probability, but  results in an increasing latency;
					% 2) increasing CTU numbers $C^t$ could improve the UE detection success probability, but it results in low resource utilization efficiency.
					% %It is necessary to optimize these parameters.

					In this work, we aim to tackle the problem of optimizing the MCG-GF-NOMA configuration defined by parameters ${\rm CG}^t_i\{N^t_{{\rm CTU},i}, N^t_{{\rm start},i}, N^t_{{\rm repe},i}\}$ for each subframe $t$.
					At each subframe $t$, the BS aims at maximizing a long-term objective $R_t$ related to the average number of UEs that have successfully send data with respect to the stochastic policy $\pi$ that maps the current observation history $O^t$ to the probabilities of selecting each possible parameters in $A^t$. 
					%Thus, the optimization problem can be formulated as
					This optimization problem (P1) can be formulated as:
					\begin{align}\label{P1}
						({\rm P1}:)&\max\limits_{\pi ({A^t}| {{O^t}} )}\quad \sum\limits_{k = t}^\infty {{\gamma ^{k - t}}}{{\mathbb E}_\pi }[N_{\rm suc}^k]\\
						s.t.\quad
						&\sum\limits_{i = 1}^{{N_{\rm CG}}} N^t_{{\rm CTU},i}  = N^t_{{\rm CTU},{\rm SCG}}, \label{st1}\\
						&N^t_{{\rm start},i}+ N^t_{{\rm repe},i}+3= N_{\rm slot}, \forall i\in[1,N_{\rm CG}], \label{st2}\\
						&N^t_{{\rm start},i}< N^t_{{\rm start},i+1}<N_{\rm slot}-3,  \forall i\in[1,N_{\rm CG}], \label{st3}
					\end{align}
					where $\gamma \in [0,1)$ is the discount factor for the performance accrued in the future subframes, and $\gamma=0$  means that the agent just concerns the immediate reward.
					The  CTU resource constraint in \eqref{st1} is set  to compare with the SCG-GF-NOMA scheme, where $N^t_{{\rm CTU},{\rm SCG}}$ is the configured CTU numbers for the SCG-GF-NOMA.
					That is to say, the  MCG-GF-NOMA configuration  uses the same frequency resources but overlap in time and have different starting points so they do not require
					the additional resources compared to the conventional SCG-GF-NOMA scheme.
					The latency constraint in \eqref{st2} is set to satisfy the latency requirement.
					That is to say, the transmission must be completed in one subframe (1 ms). Otherwise, the packet will be dropped.
					The starting slot constraint in \eqref{st3} is set to support different UL packet  arrival times.

					All these constraints yield a mixed-integer non-convex problem and, in general, there is no standard method for solving this kind of problem efficiently. Additionally, since the dynamics of the MCG-GF-NOMA system is Markovian over the continuous subframes, this is a Partially Observable Markov Decision Process (POMDP) problem that is generally intractable for the conventional convex optimization algorithms due to their limitation in overcoming the dynamic in the environment. 
					Here, partial observation refers to that a BS can not fully know all the information of the communication environment,  including, but not limited to, the channel conditions, the random collision process, and the traffic statistics.
					The search space is expanded as the number of parameters increases, which also makes the conventional gradient-based optimization techniques unsuitable. 
					%Therefore, the Reinforcement Learning (RL) algorithm, which empowers the agent to make decisions by learning from the environment, is invoked to solve the formulated problem.
					\textcolor{red}{
						The  deep  reinforcement  learning  (DRL)  is  regarded  as  powerful  tool  to  address  complex dynamic  control  problems  in POMDP. 
						The reasons in choosing DQN are that: 1) the Deep  Neural  Network  (DNN)  function approximation is able to deal with several kinds of partially observable problems \cite{sutton2018reinforcement,mnih2015human}; 2) DQN has the potential to accurately approximate the desired value function while addressing a problem with very large state spaces; 3) DQN is with high scalability, where the scale of its value function can be easily fit to a more complicated problem; 4) a variety of libraries have been established to facilitate building DNN architectures and accelerate experiments, such as TensorFlow, Pytorch, Theano, Keras, and etc..}
					The goal of deploying and designing the MCG-GF-NOMA is for maximizing the long-term benefits, which falls into the field of the DRL algorithm for the reason that this algorithm can monitor the reward resulting from its actions and incorporate farsighted system evolution instead of myopically optimizing current benefits.

					\section{Proposed Optimization Solution}

					In this section, we  propose a Cooperative Multi-Agent Double Deep Q-Network (CMA-DDQN) approach to tackle  the problem (P1), which breaks down the selection in high-dimensional action space into multiple parallel sub-tasks.

					The aim of the CMA-DDQN model is to enable the agent to carry out the optimal actions to maximize the long-term sum reward. The principle of the CMA-DDQN model is maximizing the long-term sum reward instead of aiming for maximizing the reward at a particular subframe. 
					Thus, in the CMA-DDQN model, the selected action may not be the optimal choice for the current subframe, but the optimal choice for pursing long-term benefits. 
					In this paper, the parameters configuration of MCG-GF-NOMA is considered as discrete, so the value-based RL algorithm is invoked.  
					The state space, action space, reward
					function design of the proposed CMA-DDQN based algorithm are specified.

					% In this section, we introduce a Cooperative  Multi-Agent  learning  technique  based  on  the  DDQN  (CMA-DDQN) algorithm  to tackle  the problem (P1).
					% The aim of the CMA-DDQN model is to enable the agent to carry out the optimal actions to maximize the long-term sum reward. The principle of the CMA-DDQN model is to maximize the long-term sum reward instead of aiming for maximizing the reward at a particular subframe. 
					% Thus, in the CMA-DDQN model, the selected action may not be the optimal choice for the current subframe, but the optimal choice for pursing long-term benefits. 
					% In this paper, the parameters configuration of MCG-GF-NOMA is considered as discrete, so the value-based RL algorithm is invoked.  
					% The state space, action space, reward
					% function design of the proposed CMA-DDQN based algorithm are specified.

					%\subsection{Preliminary}

					\subsection{Reinforcement Learning Framework}
					\label{RL_framework}
					To optimize the number of successfully served UEs in MCG-GF-NOMA system, we consider a RL-agent deployed at the BS to interact with the environment in order to choose appropriate actions progressively leading to the optimization goal. 
					We  define $S \in \cal S$, $A \in \cal A$, and $R \in \cal R$ as any state, action, and reward from their corresponding sets, respectively.
					At the beginning of each subframe $t$, the RL-agent first observes the current state $S^t$ corresponding to a set of previous observations ${U^{t'}}$ for all prior subframes (${t'=1,...,t-1}$) in order to select an specific action $A^t\in {\cal A}(S^t)$.
					% After carrying out actions, the agent obtains a reward $r_t$ based on the energy consumption of the UAV and the connectivity
					% condition. .
					After carrying out the action $A^t$, the RL-agent transits to a new observed state $S^{t+1}$ and obtains a corresponding reward $R^{t+1}$ as the feedback from the environment, which is designed based on the new observed state $S^{t+1}$ and guides the agent to achieve the optimization goal.
					After enough iterations, the BS can learn the optimal policy that maximizes the long-term rewards.
					% As the optimization goal is to maximize the number of the successfully served UEs under the latency constraint, we define the reward $R^{t+1}$ as 
					% \begin{align}\label{reward}
						% r^{t+1}=V_{\rm success}^t.    
						% \end{align}

					At each subframe $t$, a Q-value is calculated based on the current state and previously taken actions. Thus, the state, action and Q-value is stored in a Q-function, $Q(S^t,A^t)$, which determines the decision policy $\pi$. 
					The Q-value and Q-function are updated based on the current state, previously taken actions and the received reward by following the principle 
					
					\begin{align}\label{Qtable1}
						&Q(S^t,A^t)    \\ \nonumber
						&=Q(S^t,A^t)+\lambda[R^{t+1}+\gamma \mathop {\max }\limits_{A \in {\cal A}} Q(S^{t+1},A)-Q(S^{t},A^t) ],
					\end{align}
					The detailed descriptions of the state, action and reward of problem (P1) are introduced
					as follows.
					
					\subsubsection{States in the Q-learning Model}In terms of the state space of the proposed CMA-DDQN model, it contains five parts: the number of the collision CTUs ${N_{\rm CC}^{t'}}$, the number of the idle CTUs ${N_{\rm IC}^{t'}}$, the  number of the singleton CTUs ${N_{\rm SC}^{t'}}$,  the number of UEs  that have been successfully detected and decoded under the latency constraint ${N_{\rm suc}^{t'}}$, and the number of UEs  that have been successfully detected but not successfully decoded  ${N_{\rm fdec}^{t'}}$.

					\subsubsection{Actions in the Q-learning Model}
					Practically, the MCG-GF-NOMA system is always configured with multiple CGs to serve UEs with random traffic.
					In this section, we study the problem (P1) of optimizing the resource configuration for multiple CGs each with parameters $CG^t=\{N^t_{{\rm CTU},i}, N^t_{{\rm start},i}, N^t_{{\rm repe},i}\}_{i=1}^{N_{\rm CG}}$, where $N^t_{{\rm CTU},i}$ is chosen from the set of the number of the CTUs ${\cal N}_{{\rm CTU}}$, $N^t_{{\rm start},i}$ is chosen from the set of the value of the repetitions ${\cal N}_{{\rm start}}$, and $N^t_{{\rm repe},i}$ is chosen from the set of the value of the repetitions ${\cal N}_{{\rm repe}}$.
					%${\rm CG}^t_i\{N^t_{{\rm CTU},i}, N^t_{{\rm start},i}, N^t_{{\rm repe},i}\}$
					This joint optimization by configuring each parameter in each CG can improve the overall data transmission performance.
					%Note that each CG shares the uplink resource in the same bandwidth. 
					However, considering multiple CGs results in the increment of observations space, which exponentially increases the size of state space.
					For example, the number of available actions
					corresponds to the possible combinations of configurations
					$\left|{\cal A}\right|=\prod\limits_{i=1}^{N_{\rm CG}} {(\left|{\cal N}_{{\rm CTU},i}\right|\times \left|{\cal N}_{{\rm start},i}\right| \times \left|{\cal N}_{{\rm repe},i}\right|)}$.
					To train Q-agent with this expansion, the requirements of time and computational resources greatly increase.
					In view of this, we revise the configured parameters by considering the constraints from  \eqref{st1} to \eqref{st3}.
					%In view of this, we study Q-learning with value function approximation to design uplink resource configuration approaches for the multi-parameter scenario.
					\begin{figure}[htbp!]
						\label{Action1}
						\renewcommand{\algorithmicrequire}{\textbf{Input:}}
						\renewcommand{\algorithmicensure}{\textbf{Output:}}
						 \removelatexerror
						\begin{algorithm}[H]
							\caption{Generate the set of actions for the number of CTUs configuration.}%算法名字
							\LinesNumbered %要求显示行号
							\KwIn{Set of number of CTUs ${\cal N}_{\rm CTU}$, Length of CTUs set ${N}_{\rm CTU}=\left| {\cal N}_{{\rm CTU}} \right|$, Number of configured CTUs for the SCG-GF-NOMA $N^t_{{\rm CTU},{\rm SCG}}$, 
								Number of the configured CG at each subframe $N_{\rm CG}$.}%输入参数
							\KwOut{The set of actions for the number of CTUs configuration ${\cal  A}_{\rm CTU}^t$}%输出
							Define set ${\cal  A}_{\rm CTU}^t$;  
							
							Generate the initial index matrix:  $X \in {\mathbb C}^{1 \times N_{\rm CG}}$ with all the elements equaling to 0;

							% Generate the all one index matrix:  ${ Index_{\rm one}} \in {\mathbb C}^{1 \times N_{\rm CG}}$ with all the elements equaling to 1;\;
							
							Generate the max index  matrix:  ${ X}_{\rm max} \in {\mathbb C}^{1 \times N_{\rm CG}}$ with all the elements equaling to $N_{\rm CTU}$;

							The total searching steps  $S_{teps} =\prod\limits_{i=1}^{N_{\rm CG}} {{X}_{\rm max}[i]}$;
							
							\For {j  $\leftarrow$ 1 to  $S_{teps}$ } {
								\If {$\sum\limits_{i=1}^{N_{\rm CG}} {{\cal N}_{\rm CTU}[X[i]]}=N^t_{{\rm CTU},{\rm SCG}}$} {Put action $A_{\rm CTU}^t=\{{\cal N}_{\rm CTU}[X[i]], \forall i \in \left[1,N_{\rm CG}\right]
									\}$ into the action set ${\cal  A}_{\rm CTU}^t$;
								}

								\For{k $\leftarrow$ 1 to ${N}_{\rm CG}$} { 
									$X[-k]+=1$;
									
									\textbf{if} $X[-k] \textless X_{\rm max}[-k]:$ \textbf{break};
									
									$X[-k]\%=X_{\rm max}[-k]$.
									
								}   
							}
						\end{algorithm}
					\end{figure}
					
					First, considering the CTU resource constraint  $\sum\limits_{i = 1}^{{N_{\rm CG}}} N^t_{{\rm CTU},i}  = N^t_{{\rm CTU},{\rm SCG}}$ as presented  in \eqref{st1}, we could obtain the action set ${\cal A}_{\rm CTU}^t$, which consists of the actions $A_{\rm CTU}^t\in {\cal A}_{\rm CTU}^t$ with $A_{\rm CTU}^t=\{{N_{{\rm CTU},1}^t},...,{N_{{\rm CTU},N_{\rm CG}}^t}\}$.
					%It is worth noting that all possible combinations of the number of CTU without the  CTU resource constraint is $\prod\limits_{i=1}^{N_{\rm CG}} {\left|{\cal N}_{{\rm CTU},i}\right|}$.
					To find all possible combinations of the number of CPUs for multiple CG configurations with the CTU resource constraint, we follow the \textbf{Algorithm 1}.

					%according to the \textbf{Algorithm 1}, where $A_{\rm CTU}^t\in {\cal A}_{\rm CTU}^t$, $A_{\rm CTU}^t=\{{N_{{\rm CTU},1}^t},...,{N_{{\rm CTU},N_{\rm CG}}^t}\}$,  $N^t_{{\rm CTU},i}\in {\cal N}_{{\rm CTU}}$ and ${\cal N}_{{\rm CTU}}$ is the set of the number of CPUs.
					
					In addition, considering the starting slot constraint $N^t_{{\rm start},i}< N^t_{{\rm start},i+1} <N_{\rm slot}-3,  \forall i\in[1,N_{\rm CG}]$ in \eqref{st3}, 
					we could obtain the action set ${\cal A}_{\rm start}^t$, which consists of the actions $A_{\rm start}^t\in {\cal A}_{\rm start}^t$ with $A_{\rm start}^t=\{N_{{\rm start},1}^t,...,N_{{\rm start},N_{\rm CG}}^t\}$.
					Similarly, following the \textbf{Algorithm 1}, we can get all possible combinations of the starting slots for multiple CG configurations with the starting slot constraint.
					Different from the CTU action set, in step 6, the constraint should be starting slot constraint.

					According to the latency constraint  in \eqref{st2}, we have  $N^t_{{\rm repe},i}= N_{\rm slot}-3-N^t_{{\rm start},i}, \forall i,$. 
					Therefore, two actions set ${\cal A}_{\rm CTU}^t$ and ${\cal A}_{\rm start}^t$ is enough to characterize the multiple CG configurations defined by parameters ${\rm CG}^t_i\{N^t_{{\rm CTU},i}, N^t_{{\rm start},i}, N^t_{{\rm repe},i}\}$.
					
					%the set of actions  for starting slot configuration is

					\subsubsection{Reward Function in the Q-Learning Model}
					
					As the optimization goal is to maximize the number of the successfully served UEs under the latency constraint, we define the reward $R^{t+1}$ as 
					\begin{align}\label{reward}
						R^{t+1}=N_{\rm suc}^t,
					\end{align}
					where ${N_{\rm suc}^{t}}$ is the number of UEs  that have been successfully detected and decoded under the latency constraint.

					% \begin{align}
						%     {\cal A}_{\rm start}^t=\{A^t_{{\rm start}},A^t_{{\rm start}}=\{N^t_{{\rm start},i},
						%     0\le N^t_{{\rm start},i}\le N_{\rm slot}-4,N^t_{{\rm start},i}< N^t_{{\rm start},i+1},i=1,2,...,N_{\rm CG}\}\}
						% \end{align}

					% set of actions for CTUs number configuration is
					% \begin{align}
						%   {\cal A}_{\rm CTU}^t=\{A^t_{{\rm CTU}},A^t_{{\rm CTU}}=\{N^t_{{\rm CTU},i}, \sum\limits_{i = 1}^{{N_{\rm CG}}} N^t_{{\rm CTU},i}  = N^t_{{\rm CTU},{\rm SCG}},i=1,2,...,N_{\rm CG}\}\},
						% \end{align}
					% where $N^t_{{\rm CTU},i}\in {\cal N}^t_{{\rm CTU}}$ and ${\cal N}^t_{{\rm CTU}}$ is the set of the number of CPUs.

					%After enough iterations, the BS can learn the optimal policy that maximizes the long-term rewards.

					\subsection{Cooperative Multi-Agent DDQN Approach}

					When the number of actions and states is small, the RL algorithm can efficiently obtain the optimal policy. However, when a large number of actions and states exist, which will inevitably result in massive computation latency and severely affect the performance of the RL algorithm. 
					To address this issue, DRL is introduced, where DRL can directly control the behavior of each agent and solve complex decision-making problems, through interaction with the environment\cite{sutton2018reinforcement,mnih2015human}. 
					In addition,  Multi-Agent RL (MA-RL) is introduced with centralized or decentralized rewards. 
					In MA-RL with centralized rewards, all agents receive a common (central) reward, while in MA-RL with decentralized rewards, every agent obtains a distinct reward \cite{9084325}. 
					However, in  MA-RL with decentralized rewards, all agents may compete with each other, i.e., agents may act in a selfish behavior for requiring the highest reward which may affect the global network performance. 
					To convert this selfishness into cooperative behavior, the same reward may be assigned to all agents \cite{8792382}. In this section, we apply the Cooperative Multi-Agent technique based DDQN (CMA-DDQN) to prevent the selfish behavior of agents.
					%Each DDQN agent controls their own action variable, and receives a common reward to guarantee the objective in (P1) cooperatively.

					%In this section,  we  propose a cooperative multi-agent DDQN (CMA-DDQN) approach to optimize the configuration of MCG-GF-NOMA, which breaks down the selection in high-dimensional action space into multiple parallel sub-tasks.Each DDQN agent controls their own action variable, and receives a common reward to guarantee the objective in (P1) cooperatively. 

					The challenge of this approach is how to evaluate each
					action according to the common reward function. For each
					DQN agent, the received reward is corrupted by massive noise, where its own effect on the reward is deeply hidden in the effects of all other DQN agents. 
					For instance, a positive action can receive a mismatched low reward due to other DQN agents’ negative actions. 
					Fortunately, in our scenario, all DQN agents are centralized at the BS, which means that all DQN agents can have full information among each other.
					The CMA-DDQN algorithm utilizes the experience replay technique to enhance the convergence performance of RL. When updating the  CMA-DDQN algorithm, mini-batch samples are selected randomly from the experience memory as the input of the neural network,  which  breaks  down  the  correlation  among  the  training samples.  
					In  addition,  through averaging  the  selected  samples,  the  distribution  of  training  samples  can  be  smoothed,  which avoids the training divergence.
					We define $A^t_x$ as the action selected by the $x$th
					agent.  Each $x$th agent is responsible for updating  the value  $Q(S^t,A^t_x)$ of action $A^t_x$ in state $S^t$, where the state variable ${{S ^{t}}} = [A^{t-1}, U^{t-1}, A^{t-2}, U^{t-2}, ..., A^{t-M_{o}}, U^{t-M_{o}}]$ only includes information about the last $M_o$ RTTs.
					All agents receive the same reward $R^{t+1}$  at the end of each subframe.

					%Suppose that the set of the number of CPUs is ${\cal N}^t_{{\rm CTU}}$, i.e., $N^t_{{\rm CTU},i}\in {\cal N}^t_{{\rm CTU}}$, 

					%Thus, the set of action $A_1^t=\{N^t_{{\rm CTU},i},i\in [1,N_{\rm CG}]\}$, where $N^t_{{\rm CTU},i}\in $The set of $N^t_{{\rm CTU},i}$ value is {8,16,32,40}

					\begin{figure}[htbp!]
						\label{DQN}
						\renewcommand{\algorithmicrequire}{\textbf{Input:}}
						\renewcommand{\algorithmicensure}{\textbf{Output:}}
						 \removelatexerror
						\begin{algorithm}[H]
							\caption{CMA-DQN Based MCG-GF-NOMA Uplink Resource Configuration}%算法名字
							\LinesNumbered %要求显示行号
							\KwIn{: Action space $\cal A$ and Operation Iteration I.}%输入参数
							% \KwOut{output result}%输出
							Algorithm hyperparameters: learning rate $\lambda_{RMS} \in (0, 1])$, discount rate $\gamma \in [0, 1) $,  $\epsilon$-greedy rate $\epsilon \in (0, 1]$,  target network update frequency $Y$; 
							
							Initialization of replay memory $M$ to capacity $D$, the state-action value function $Q(S,A,{\bm\theta})$,
							the parameters of primary Q-network $\bm\theta$, and the target Q-network $\bm{\bar \theta}$;
							
							\For{Iteration $\leftarrow$ 1 to I}{
								Initialization of $S^1$ by executing a random action $A^0_x$;
								
								\For {t $\leftarrow$ 1 to T}{
									% Update $\mu^0$ using Eq. (2);\;
									
									\textbf{if} {$p_\epsilon <\epsilon$} \textbf{Then} select a random action $A^t_x$ from ${\cal A}_x$
									
									\textbf{else} select
									${A^t_x} = \mathop {\arg \max }\limits_{a \in {\cal A}_x} Q({S^t},A_x^t,\bm\theta_x )$.
									
									The BS broadcasts $A^t_x$ and backlogged UEs attempt communication in the $t$th subframe;
									
									The BS observes state $S^{t+1}$, and calculate the related reward $R^{t+1}$;
									
									Store transition $(S^t, A^t_x, R^{t+1}, S^{t+1})$ in replay memory $M_x$;
									
									Sample random minibatch of transitions $(S^t, A^t_x, R^{t+1}, S^{t+1})$ from replay memory $M_x$;
									
									Perform a gradient descent step and update parameters $\bm\theta_x$ for $Q(S^t,A_x^t,{\bm\theta_x})$ using \eqref{approximator};
									
									Update the parameter $\bm{\bar\theta}=\bm{\theta}$ of the target Q-network every $Y$ steps.
								}
							}
						\end{algorithm}
					\end{figure}

					The DDQN agents are trained in parallel.
					Each agent $x$ parameterizes the action-state value function $Q(S^t,A^t_x)$ by using a function $Q(S^t,A^t_x,\bm{\theta}_x)$, where $\bm{\theta}_x$ represents the weights matrix of a multiple layers DNN  with fully-connected layers.
					The variables in the state $S^t$ is fed in to the DNN as the input; the Rectifier Linear Units (ReLUs) are adopted as intermediate hidden layers; while the output layer is consisted of linear units, which are in one-to-one correspondence with all available actions in $\cal A$.
					The online update of weights matrix $\bm{\theta}_x$ is carried out along each training episode by using DDQN\cite{van2015deep}. 
					Accordingly, learning takes place over multiple training episodes, where each episode consists of several RTT periods. In each RTT, the parameters $\bm{\theta}_x$ of the Q-function
					approximator $Q(S^t,A^t_x,\bm{\theta}_x)$ are updated using RMSProp optimizer\cite{tieleman2012lecture} as
					\begin{align}\label{RMS}
						{\bm{\theta} ^{t + 1}_x} = {\bm{\theta} ^{t}_x} - {\lambda _{\mathrm{RMS}}}\nabla {L_x^{\mathrm{DDQN}}}({\bm{\theta} ^t_x}) 
					\end{align}
					where $\lambda _{\mathrm{RMS}} \in (0,1]$ is RMSProp learning rate, $\nabla {L_x^{\mathrm{DDQN}}}({\bm{\theta}^t_x})$ is the
					gradient of the loss function ${L_x^{\mathrm{DDQN}}}({\bm{\theta} ^t_x})$ used to train the state-action value function. The gradient of the loss function is defined as
					\begin{align}\label{approximator}
						&\nabla {L_x^{\mathrm{DDQN}}}({\bm{\theta}^t_x})\nonumber\\
						&= {{\rm E}_{{S^j},{A^j_x},{R^{j + 1}},{S^{j + 1}}}}[({R^{j + 1}} + \gamma \mathop {\mathop {\max }\limits_{a \in {\cal A}} \ Q({S^{j + 1}}, {A_x^j},{{\bar {\bm{\theta}} }^t_x})} \\ \nonumber  
						& - Q({S^j},{A^j_x},{\bm{\theta} ^t_x})){\nabla _{\bm{\theta}_x }}Q({S^j},{A^j_x},{\bm{\theta} ^t_x})],
					\end{align}
					where the expectation is taken over the minibatch, which are randomly selected from previous samples $(S^j, A^j_x, S^{j+1}, R^{j+1})$ 
					for $j \in \{t-M_r,...,t\}$ with $M_r$ being the replay memory size\cite{sutton2018reinforcement}. 
					When $t-M_r$ is negative, it represents to include samples from the previous episode. Furthermore, $\bar {{\bm{\theta}}^t}$  is the target Q-network in DDQN that is used to estimate the future value of the Q-function
					in the update rule, and $\bar {\bm{\theta}}^t$ is periodically copied
					from the current value $ {\bm{\theta}}^t$ and kept unchanged for several episodes.

					Through calculating the expectation of the selected
					previous samples in minibatch and updating the ${\bm{\theta} ^{t}}$ by \eqref{RMS}, the DDQN value function $Q(s,a,\bm{\theta})$ can be obtained. The detailed CMA-DDQN algorithm is presented in \textbf{Algorithm 2.}
					We consider $\epsilon$-greedy approach to
					balance exploitation and exploration in the actor of the Q-Agent, where  $\epsilon$
					is a positive real number and $\epsilon<1$. In each subframe $t$, the Q-agent randomly generates a probability $P_\epsilon^t$to
					compare with $\epsilon$.
					Then, with the probability $\epsilon$, the algorithm randomly chooses an action from the remaining feasible actions to improve the estimate of the non-greedy action’s
					value. 
					With the probability $1-\epsilon$, the algorithm exploits the current knowledge of the Q-value table to choose the action
					that maximizes the expected reward.
					
					%The detailed DQN algorithm is presented in \textbf{Algorithm 1}.

					%Through calculating the expectation of the selectedprevious samples in minibatch and updating the ${\bm{\theta} ^{t}}$ by \eqref{RMS}, the DQN value function $Q(s,a,\bm{\theta})$ can be obtained. The detailed DQN algorithm is presented in \textbf{Algorithm 1.}
					% \begin{figure}[htbp!]
						% 	\centeringq
						% 	\includegraphics[width=5.9in,height=3.4in]{DQN.pdf}
						% 	\caption{The CMA-DQN agents and environment interaction in the POMDP.}
						% 	\label{fig:9}
						% \end{figure}
					%At the beginning of each subframe $t$, the decision is made by the BS according to the transmission receptions ${U^{t'}}$ for all prior subframes (${t'=1,...,t-1}$), consisting of the  following variables: the number of the collision CTUs ${V_{\rm CC}^{t'}}$, the number of the idle CTUs ${V_{\rm IC}^{t'}}$, the  number of the singleton CTUs ${V_{\rm SC}^{t'}}$,  the number of UEs  that have been successfully detected and decoded under the latency constraint ${V_{\rm success}^{t'}}$, and the number of UEs  that have been successfully detected but not successfully decoded  ${V_{\rm faildeco}^{t'}}$.
					%We denote ${{H^t}=\{O^1, O^2,..., O^{t-1}\}}$ with ${{O^{t-1}}=\{U^{t-1}, A^{t-1}\}}$ as the observation in each subframe $t$ including  histories of all such measurements and past actions.

					\textcolor{red}{
						\subsection{Computational Complexity}
						The approximate complexity of generating the set of actions for $X$ agents is $O(XS_{step}N_{CG})$, where $S_{step}$ represents the maximum iteration steps  and $N_{CG}$ represents the element number checking and correcting. 
						The training complexity for $X$ agents, one minibatch of $I$ episodes with $T$ time-steps until
						convergence results in computational complexity is of order $O(X^2S_{step}N_{CG}IT)$ in training phase.
						The structures of the value function approximator can also be specifically designed for RL agents with sub-tasks of significantly different complexity. However, there is no such requirement in our problem, so it will not be considered.
						DNN is a better value function approximator due to its efficiency and capability in solving high complexity problems.
						%The computational complexity of the proposed CMA-DQN algorithm is of order $O(X^2S_{step}N_{CG}IT)$, where $I$ is the number of episodes and $T$ is the time-steps.
					}

					\section{Simulation Results}
					In this section, we examine the effectiveness of our proposed MCG-GF-NOMA system with CMA-DDQN algorithm
					via simulation.
					We adopt the standard network parameters listed in Table II following\cite{2Tel2018}, and
					hyperparameters for the DQN learning algorithm are listed in Table III.
					\textcolor{red}{Without loss of generality, in the simulation, we focus on the mini-slots of $N_{\rm sym}=7$ OFDM symbols for transmissions using 60 kHz ($\mu=2$) SCS, which is in line with the main guidelines for 3GPP NR performance evaluations presented in \cite{2Tel2018}.}
					\begin{table}[htbp!]
						\centering
						\caption{Simulation Parameters}
						{\renewcommand{\arraystretch}{1.2}
							%\renewcommand{\tabcolsep}{0.15cm}
							%	\begin{tabular}{|c|c|c|c|}
								{  	\rowcolors{0}{gray!25}{white}
									\begin{tabular}{|p{3.5cm}|p{3.5cm}|} 
										\hline			
										Parameters  & Value  \\ \hline 
										Numerology  factor  $\mu$ & 2  \\ 	\hline
										Number of OFDM symbols in a slot $N_{\rm sym}$ & 7 \\ 	\hline
										Path-loss exponent $\eta$ & 4  \\ \hline
										Noise power $\sigma^2$ & -132 dBm \\ \hline
										Transmission power $P$ & 23 dBm   \\ \hline
										The received SINR threshold $\gamma_{th}$ & -10 dB  \\ \hline
										Duration of traffic $T$  &  1000 ms \\ 	\hline 
										The number of the configured CTUs for the SCG-GF-NOMA  $N_{{\rm CTU},{\rm SCG}}$ & 64\\ \hline
										The set of the number of CTUs ${\cal N}_{CTU}$& $\{8, 16, 24, 32, 40,48,56\}$ \\ \hline 
										The set of the starting slot ${\cal N}_{start}$ & $\{0, 1, 2, 3, 4 \}$ \\  \hline
										% 		Bursty traffic parameter Beta($\alpha$,$\beta$) &  (3, 4)  \\ 
										% 		\hline 
										The number of static UEs for low (high) traffic $N_{\rm UE}$  &  10000 (50000) \\ 
										\hline
										%The number of bursty UEs for high traffic $N_{\rm UE}$  &  50000 \\ 
										%\hline   
										The number of time-frequency RBs $F$ & 4\\
										\hline
										Cell radius $R$ &  10 km \\ 
										\hline
										The number of slots within a subframe $N_{\rm slot}$ & 8 \\ 
										\hline
									\end{tabular}
								}
							}
							\label{table:2}
						\end{table}
						\begin{table}[htbp!]
							\centering
							\caption{Learning Hyperparameters}
							{\renewcommand{\arraystretch}{1.2}
								%\renewcommand{\tabcolsep}{0.15cm}
								%	\begin{tabular}{|c|c|c|c|}
									{	\rowcolors{0}{gray!25}{white}
										\begin{tabular}{|p{4.5cm}|p{2.5cm}|}
											\hline
											Hyperparameters  & Value  \\ \hline		 
											Learning rate  $\lambda_{RMS}$   &0.0001 \\ \hline
											Minimum exploration rate $\epsilon$ & 0.1 \\ 
											\hline
											Discount rate  $\gamma$ & 0.5   \\ \hline
											Minibatch size & 32  \\ \hline
											Replay Memory  &  10000 \\ \hline
											Target Q-network update frequency & 1000\\ 
											\hline
										\end{tabular}
									}
								}
								\label{table:3}
							\end{table}
							
							All testing performance results are obtained by averaging over 1000 episodes.
							The BS is located at the center of a circular area with a 10 km radius, and the UEs are randomly located within the cell. 
							The DQN is set with two hidden layers, each with 128 ReLU units. 
							In the following, we present our simulation
							results of multiple CG configurations in MCG-GF-NOMA system.

							% \begin{figure}[htbp!]
								% 	\centering
								% 	\includegraphics[width=3.5in,height=2.6in]{realtraffic.eps}
								% 	\caption{The real-time traffic load.
									% 	}
								% 	\label{fig:6}
								% \end{figure}

							\begin{figure}[htbp!]
								\begin{center}
									%  \begin{minipage}[t]{0.49\textwidth}
										\centering
										\includegraphics[width=3.5in,height=2.6in]{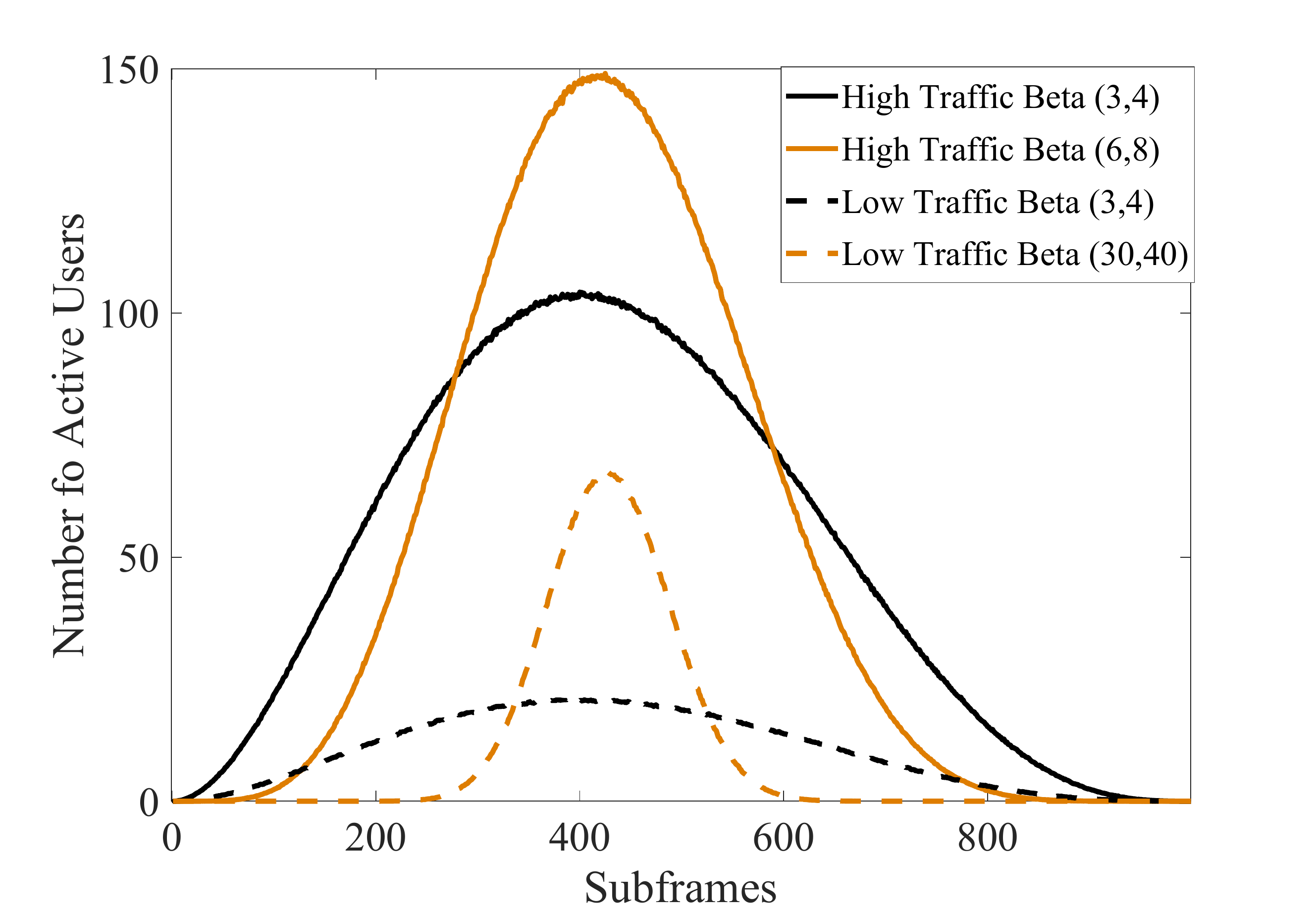}
										\vspace*{-0.4cm}
										\caption{\textcolor{red}{The real-time traffic load.}
										}
										\label{fig:6}
										%  \end{minipage}
								\end{center}
							\end{figure}

							Throughout epoch, each UE has a bursty
							traffic profile (i.e., the time limited Beta profile defined in \eqref{beta} with parameters (3, 4)\textcolor{red}{, (6, 8) or (30, 40))} that has a peak around the 400th subframe. 
							The resulting average number of newly generated
							packets is shown in Fig.~\ref{fig:6}, where the  dashed line represents the low traffic (LOW) and the solid line represents the high traffic (HIGH).

							\begin{figure}[htbp!]
								\begin{center}
									%\begin{minipage}[t]{0.49\textwidth}
									\centering
									\includegraphics[width=3.5in,height=2.6in]{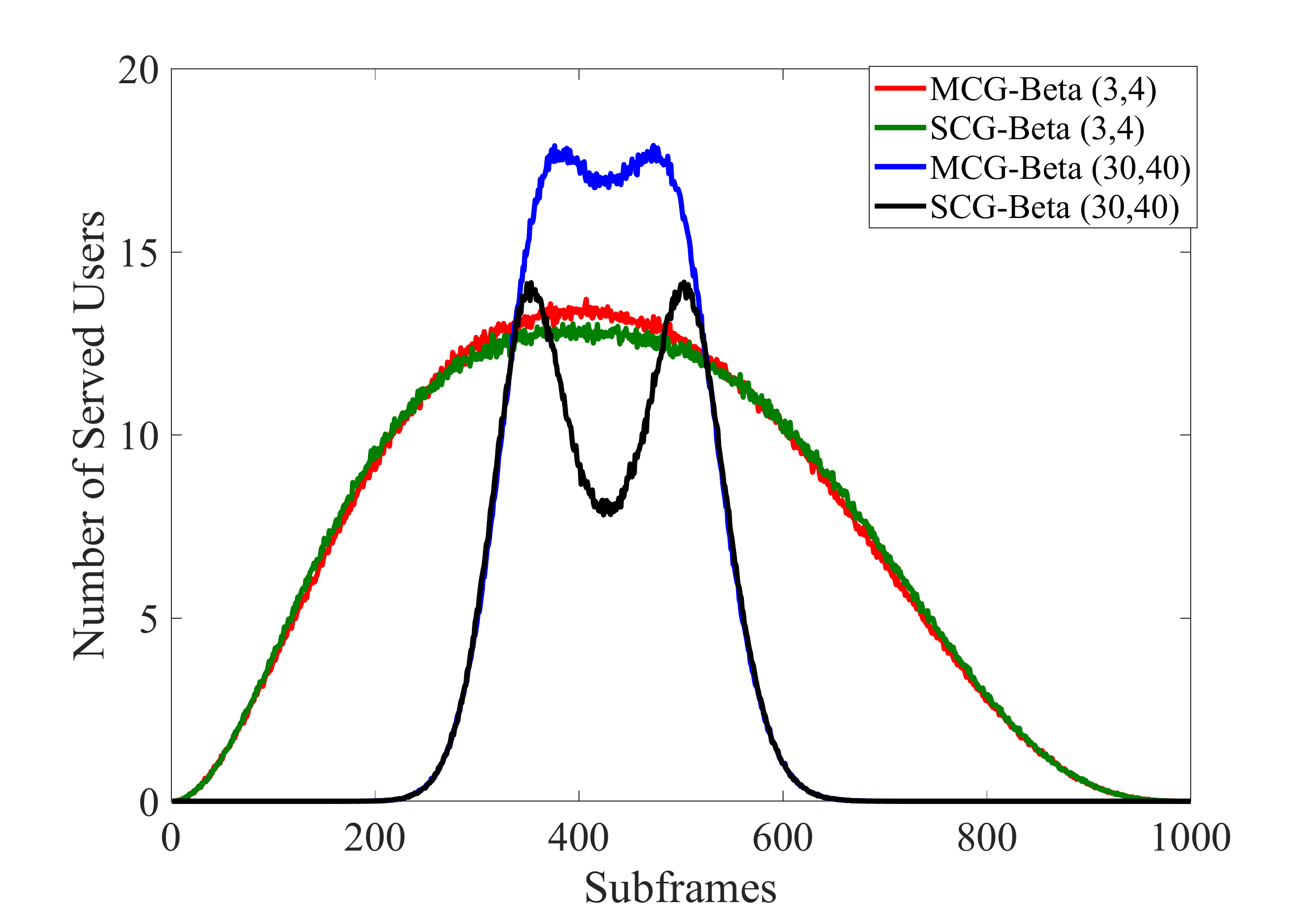}
									\vspace*{-0.4cm}
									\caption{\textcolor{red}{Average number of successfully served users in low traffic scenario.}}
									\label{fig:10}
									% \end{minipage}
							\end{center}
						\end{figure}

						\begin{figure}[htbp!]
							\begin{center}
								%\begin{minipage}[t]{0.49\textwidth}
								\centering
								\includegraphics[width=3.5in,height=2.6in]{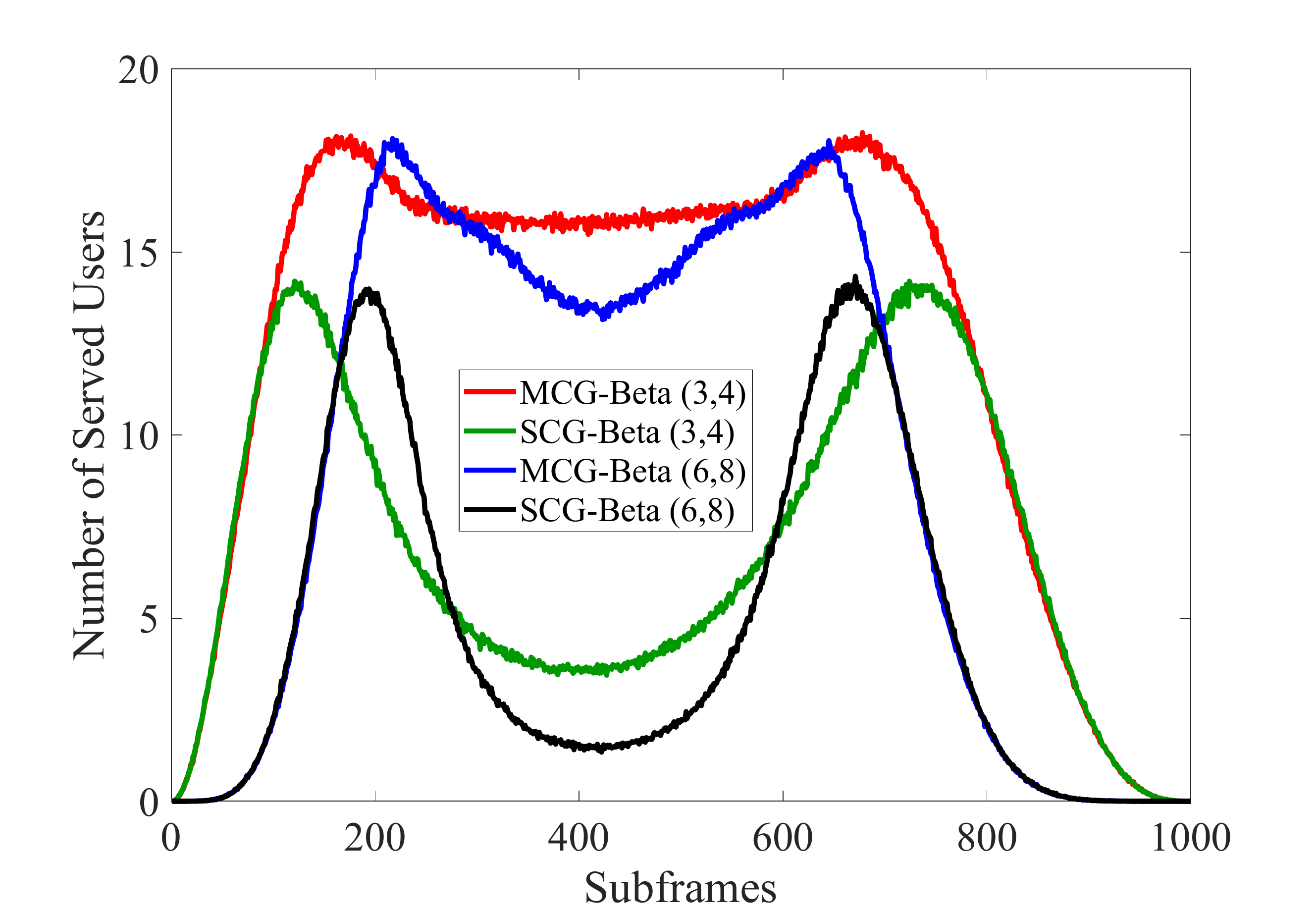}
								\vspace*{-0.4cm}
								\caption{\textcolor{red}{Average number of successfully served users in high traffic scenario.}}
								\label{fig:10-1}
								% \end{minipage}
						\end{center}
					\end{figure}

					\textcolor{red}{Fig.~\ref{fig:10} compares the number of successfully served UEs for MCG-GF-NOMA and SCG-GF-NOMA systems in low traffic scenario with parameters Beta(3, 4) and Beta(30, 40), respectively.
						Unless otherwise stated, we consider $N_{\rm CG}=5$ for the MCG-GF-NOMA system.
						It is obvious that the MCG-GF-NOMA  can increase the successfully served UEs compared with the SCG-GF-NOMA, especially for the high bursty traffic peak (Beta(30, 40)), i.e., massive access simultaneously.
						Particularly, at the peak traffic, the number of successfully served UEs in the MCG-GF-NOMA system is circa two times more than that in the SCG-GF-NOMA system.
						However, when the bursty traffic is lower (Beta(3, 4)), this advantage of MCG is not obvious.
						This  indicates  that  the MCG solution can  ensure the massive access performance of GF-NOMA in a massive URLLC scenario.}

					\textcolor{red}{Fig.~\ref{fig:10-1} compares the number of successfully served UEs for MCG-GF-NOMA and SCG-GF-NOMA systems in high traffic scenario with parameters Beta(3,4) and Beta(6,8), respectively.
						% Unless otherwise stated, we consider $N_{\rm CG}=5$ for the MCG-GF-NOMA system.
						We observe that at the peak traffic with parameter (3, 4), the number of successfully served UEs in the MCG-GF-NOMA system is circa four times more than that in the SCG-GF-NOMA system, while at the peak traffic with parameter (6, 8), the number of successfully served UEs in the MCG-GF-NOMA system is circa seven times more than that in the SCG-GF-NOMA system.
						This is in line with Fig.~\ref{fig:10} that the MCG-GF-NOMA outperform the SCG-GF-NOMA for massive access scenario.
						It should be noted that the number of successfully served UEs for MCG-GF-NOMA with Beta (6, 8) decreases slightly at the peak traffic compared with that for MCG-GF-NOMA with Beta (3, 4).
						It indicates that with ever-increasing traffic, the ability of MCG-GF-NOMA will be limited, more efficient solution should be designed.}
					
					\begin{figure}[htbp!]
						\begin{center}
							%  \begin{minipage}[t]{0.5\textwidth}
								\centering
								\includegraphics[width=3.5in,height=2.6in]{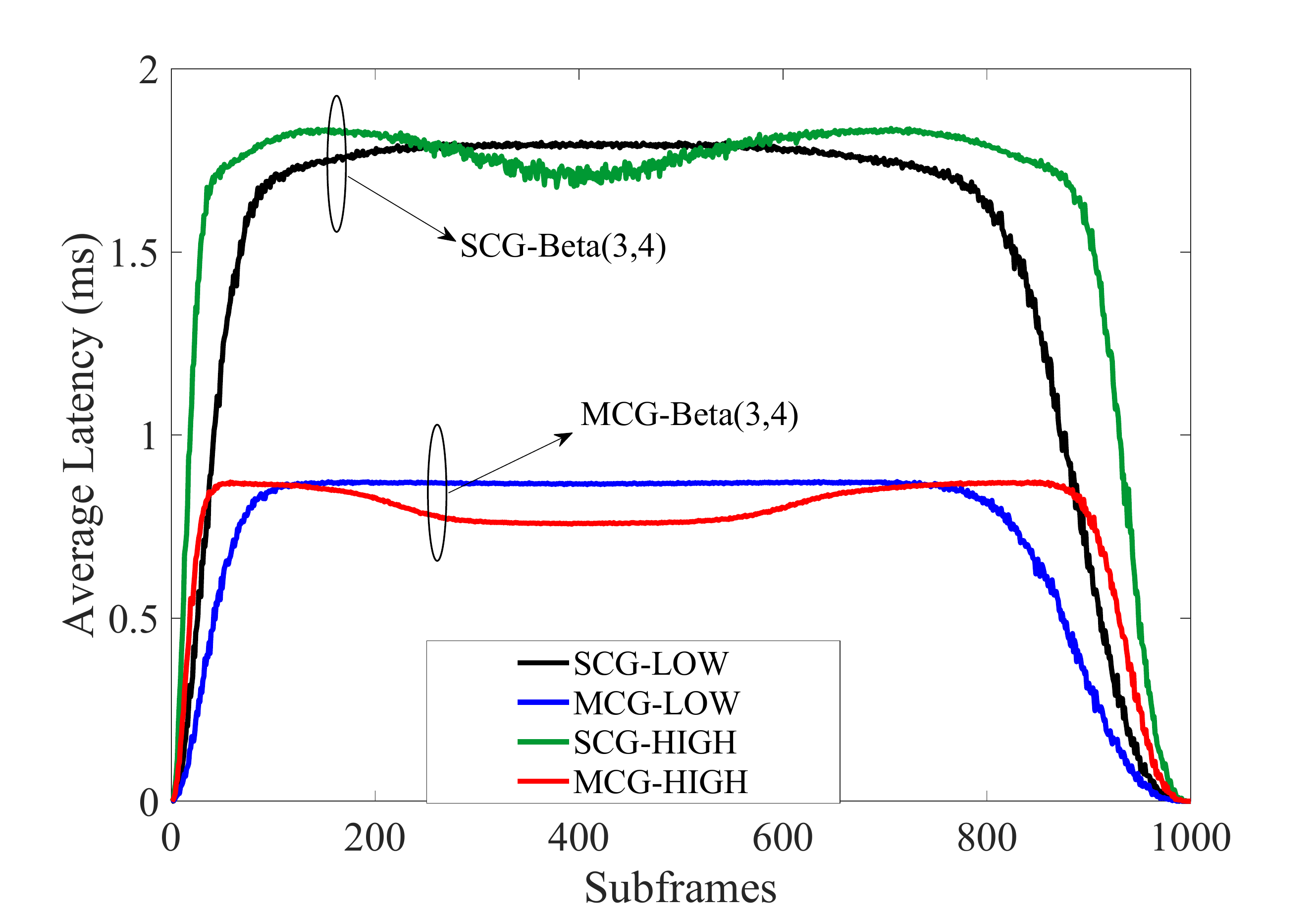}
								\vspace*{-0.4cm}
								\caption{Average latency of successfully served users.}
								\label{fig:11}
								% \end{minipage}
						\end{center}
					\end{figure}
					
					Fig.~\ref{fig:11} compares the average latency of successfully served UEs in MCG-GF-NOMA and SCG-GF-NOMA systems with both high traffic and low scenarios with parameters Beta(3, 4), respectively.
					It is obvious that the MCG-GF-NOMA can decrease the average latency of successfully served UEs compared to the SCG-GF-NOMA, for both the high traffic and low traffic scenarios.
					In particular, the MCG-GF-NOMA system could almost decrease the latency by half compared with that in the SCG-GF-NOMA system.
					This  indicates  that  the MCG solution can  ensure the low latency performance of GF-NOMA in a massive URLLC scenario.

					\begin{figure}[htbp!]
						\begin{center}
							%  \begin{minipage}[t]{0.49\textwidth}
								\centering
								\includegraphics[width=3.5in,height=2.6in]{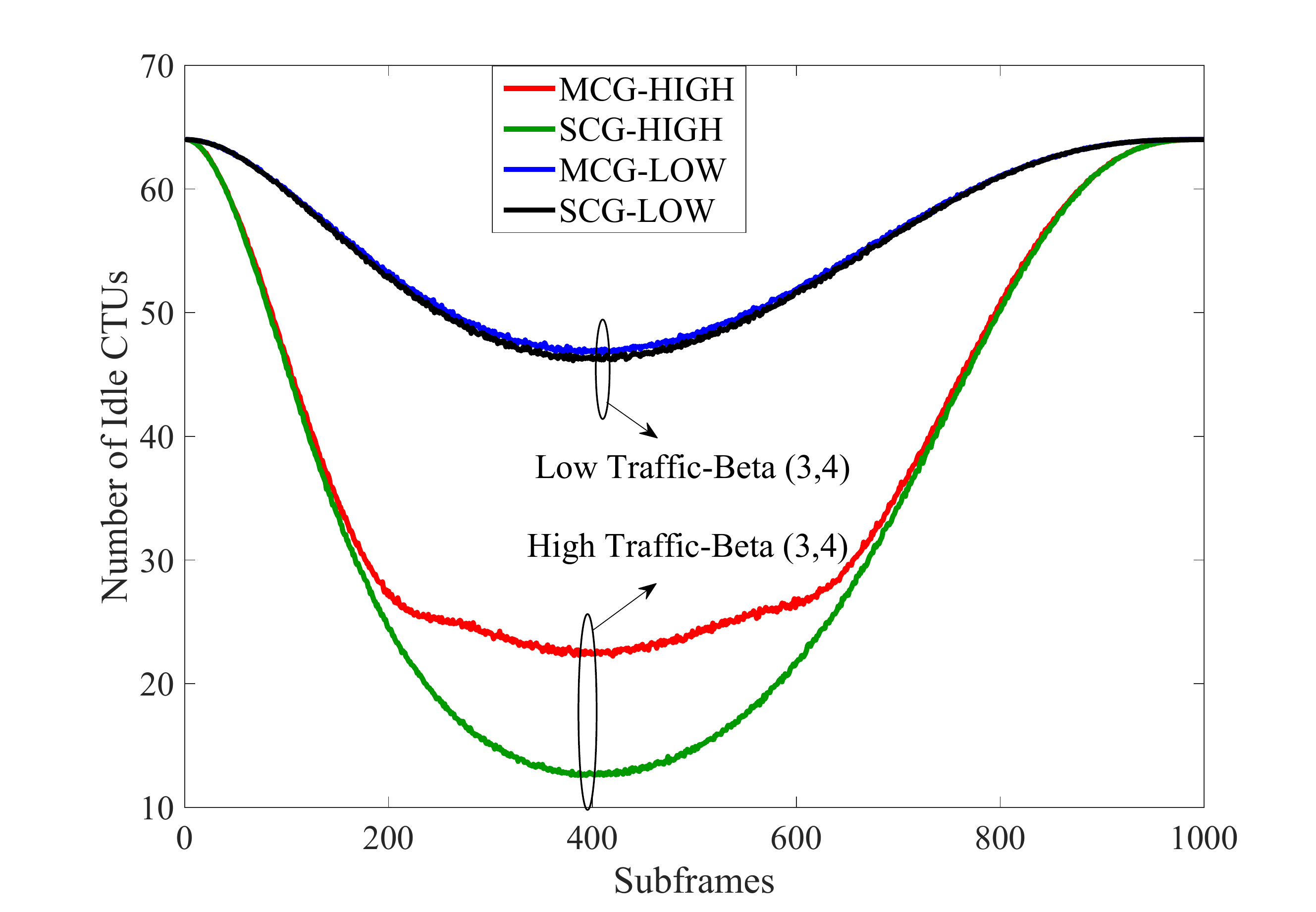}
								\vspace*{-0.4cm}
								\caption{Average number of idle CTUs.}
								\label{fig:12}
								%    \end{minipage}
						\end{center}
					\end{figure}

\begin{figure}[htbp!]
						\begin{center}
							%  \begin{minipage}[t]{0.5\textwidth}
								\centering
								\includegraphics[width=3.5in,height=2.6in]{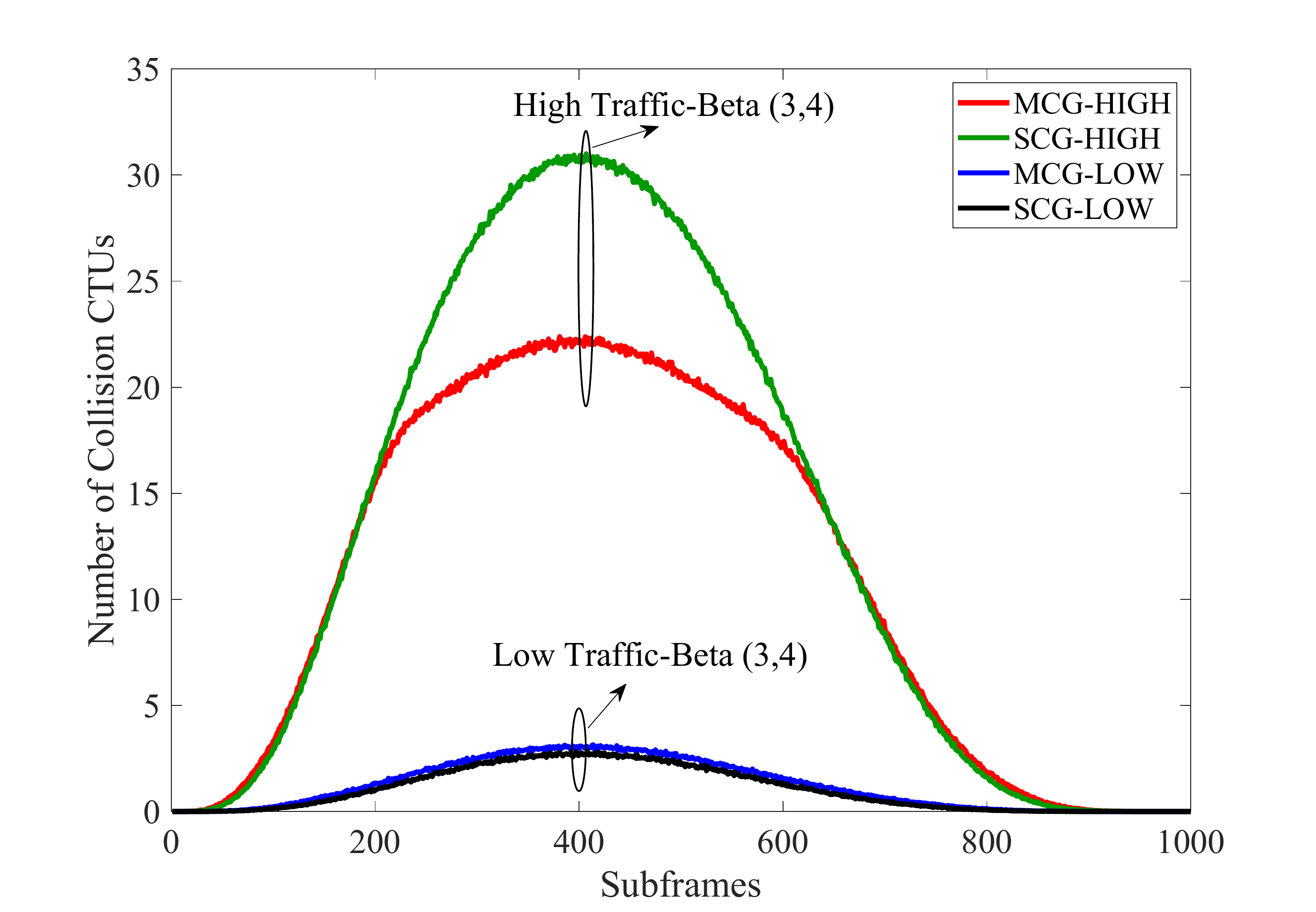}
								\vspace*{-0.4cm}
								\caption{Average number of collision CTUs.}
								\label{fig:13}
								%   \end{minipage}
						\end{center}
					\end{figure}

					Fig.~\ref{fig:12} and Fig.~\ref{fig:13} compare the average number of idle and collision CTUs in MCG-GF-NOMA and SCG-GF-NOMA systems with both high traffic and low traffic scenarios with parameters Beta(3, 4),  respectively.
					Combining with Fig.~\ref{fig:10}-Fig.~\ref{fig:11}, we observe that the multiple CGs solution can obtain better reliability and latency performance of  MCG-GF-NOMA only by using smaller CTU resources than the SCG-GF-NOMA with the single CG, especially for the high traffic scenario.
					This is due to the fact that the MCG solution mitigates the heavy traffic backlog in the SCG-GF-NOMA system, where multiple  UEs are active after the starting slot offset of one  CG will wait for the next CG period to transmit the packet.
					Consequently, the collision events are mitigated in the MCG-GF-NOMA system.
					
					\begin{figure}[htbp!]
						\begin{center}
							% \begin{minipage}[t]{0.49\textwidth}
								\centering
								\includegraphics[width=3.5in,height=2.6in]{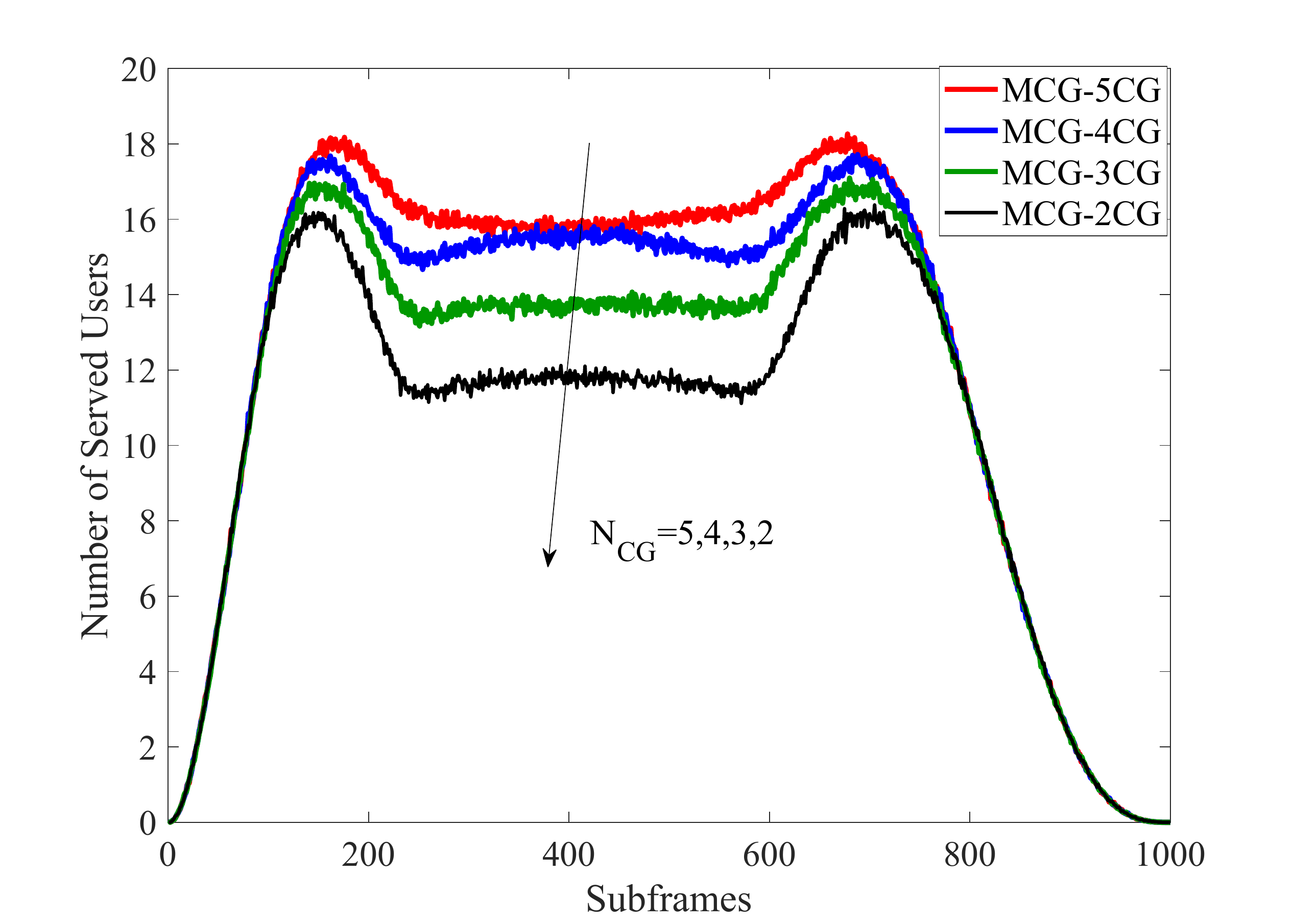}
								\vspace*{-0.4cm}
								\caption{Average number of successfully served users in MCG-GF-NOMA with different numbers of configured-grants $N_{\rm CG}$.}
								\label{fig:14}
								%   \end{minipage}
						\end{center}
					\end{figure}

\begin{figure}[htbp!]
						\begin{center}
							%     \begin{minipage}[t]{0.5\textwidth}
								\centering
								\includegraphics[width=3.5in,height=2.6in]{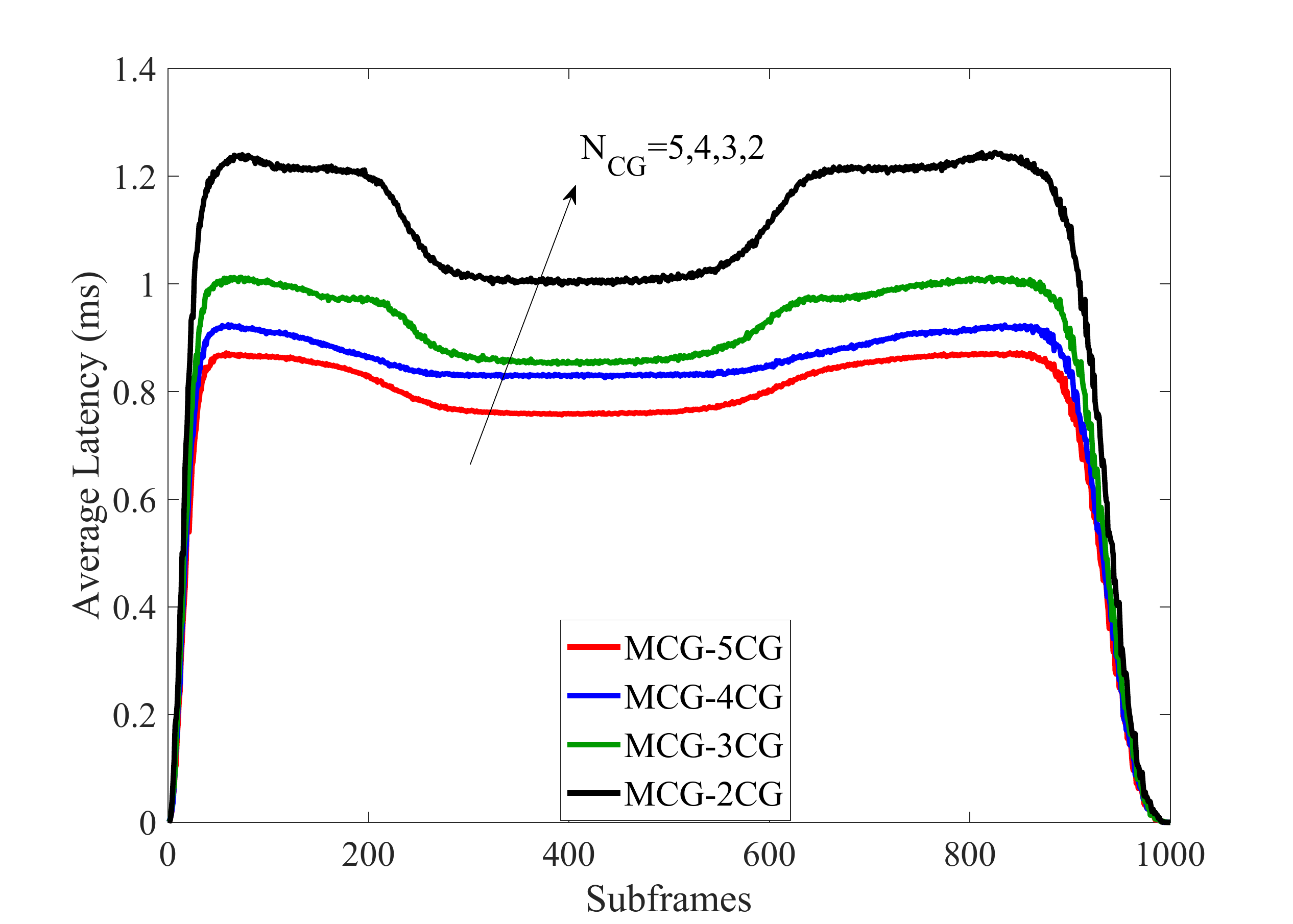}
								\vspace*{-0.4cm}
								\caption{Average latency of successfully served users in MCG-GF-NOMA with different numbers of configured-grants $N_{\rm CG}$.}
								\label{fig:15}
								%    \end{minipage}
						\end{center}
					\end{figure}

					Fig.~\ref{fig:14} and Fig.~\ref{fig:15} compare the average number of successfully served users and the average latency of successfully served users in the  MCG-GF-NOMA system with high traffic for different numbers of CGs $N_{\rm CG}$,  respectively.
					\textcolor{red}{Unless otherwise stated, we consider bursty traffic parameter Beta(3, 4) for the MCG-GF-NOMA system.}
					We observe that the average number of successfully served users increases, whereas the average latency of successfully served users decreases,  with increasing the numbers of CGs $N_{\rm CG}$. 
					The increased degree of the average number of successfully served users and the decreased degree of the average latency of successfully served users is largest at the peak traffic around the 400th subframe.
					This indicates that more CGs can improve the massive access performance of GF-NOMA in high traffic regions, which is in line of the descriptions of MCG-GF-NOMA in Section I.A.
					The MCG-GF-NOMA system could mitigate the collision events when multiple UEs are active and waiting for the CG period to transmit the packet.
					It should be noted that both the increased degree of the average number of successfully served users and the decreased degree of the average latency of successfully served users decrease with increasing the numbers of CGs $N_{\rm CG}$.
					
					%That is to say, there is an upper bound of the average number of successfully served users and a lower bound of the average latency of successfully served users in the MCG-GF-NOMA system.

					\begin{figure}[htbp!]
						\begin{center}
							%  \begin{minipage}[t]{0.5\textwidth}
								\centering
								\includegraphics[width=3.5in,height=2.6in]{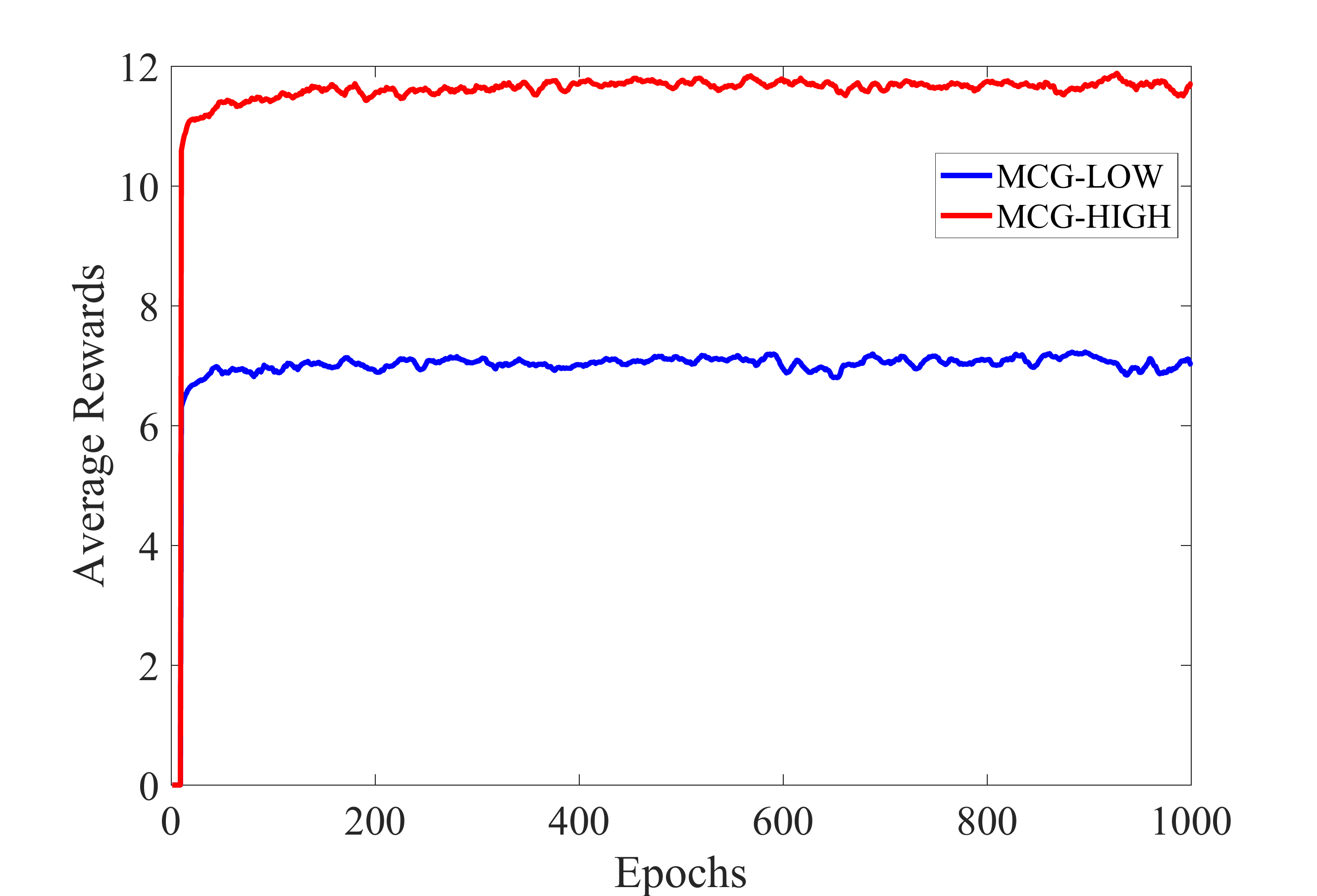}
								\vspace*{-0.4cm}
								\caption{Average received reward.}
								\label{fig:7}
								%  \end{minipage}
						\end{center}
					\end{figure}
					
					In Fig.~\ref{fig:7}, we show the system convergence process of the proposed CMA-DDQN aided MCG-GF-NOMA schemes by plotting the average reward.
					It can be intuitively seen that the proposed framework has a fast convergence speed and the episode required for system convergence is very small.

					\begin{figure}[htbp!]
						\begin{center}
							%  \begin{minipage}[t]{0.49\textwidth}
								\centering
								\includegraphics[width=3.5in,height=2.6in]{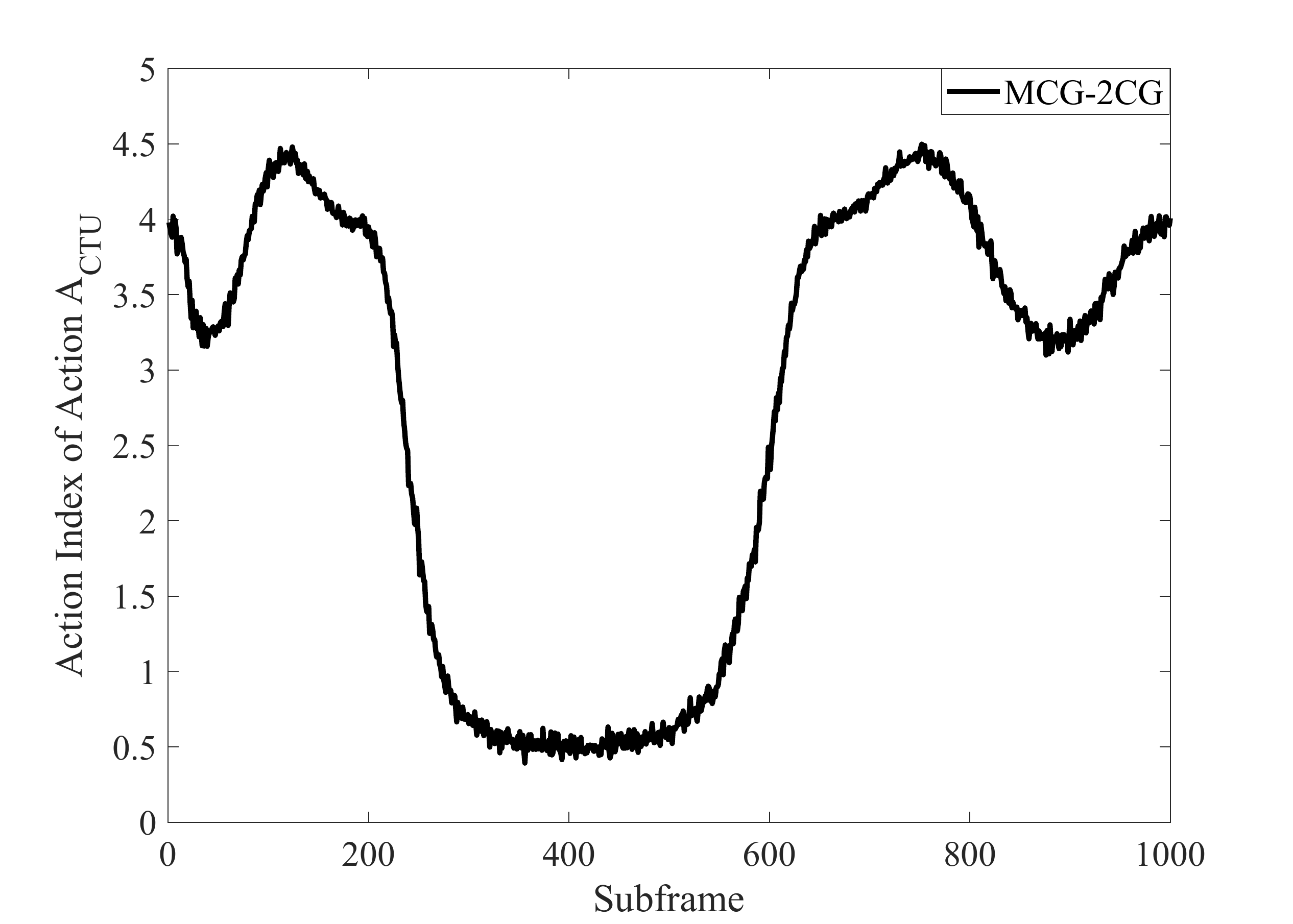}
								\vspace*{-0.4cm}
								\caption{Action index of the action $A_{\rm CTU}$ in the action set ${\cal A}_{\rm CTU}$ for MCG-GF-NOMA with $N_{\rm CG}=2.$}
								\label{fig:16}
								%   \end{minipage}
						\end{center}
					\end{figure}

\begin{figure}[htbp!]
						\begin{center}
							%  \begin{minipage}[t]{0.5\textwidth}
								\centering
								\includegraphics[width=3.5in,height=2.6in]{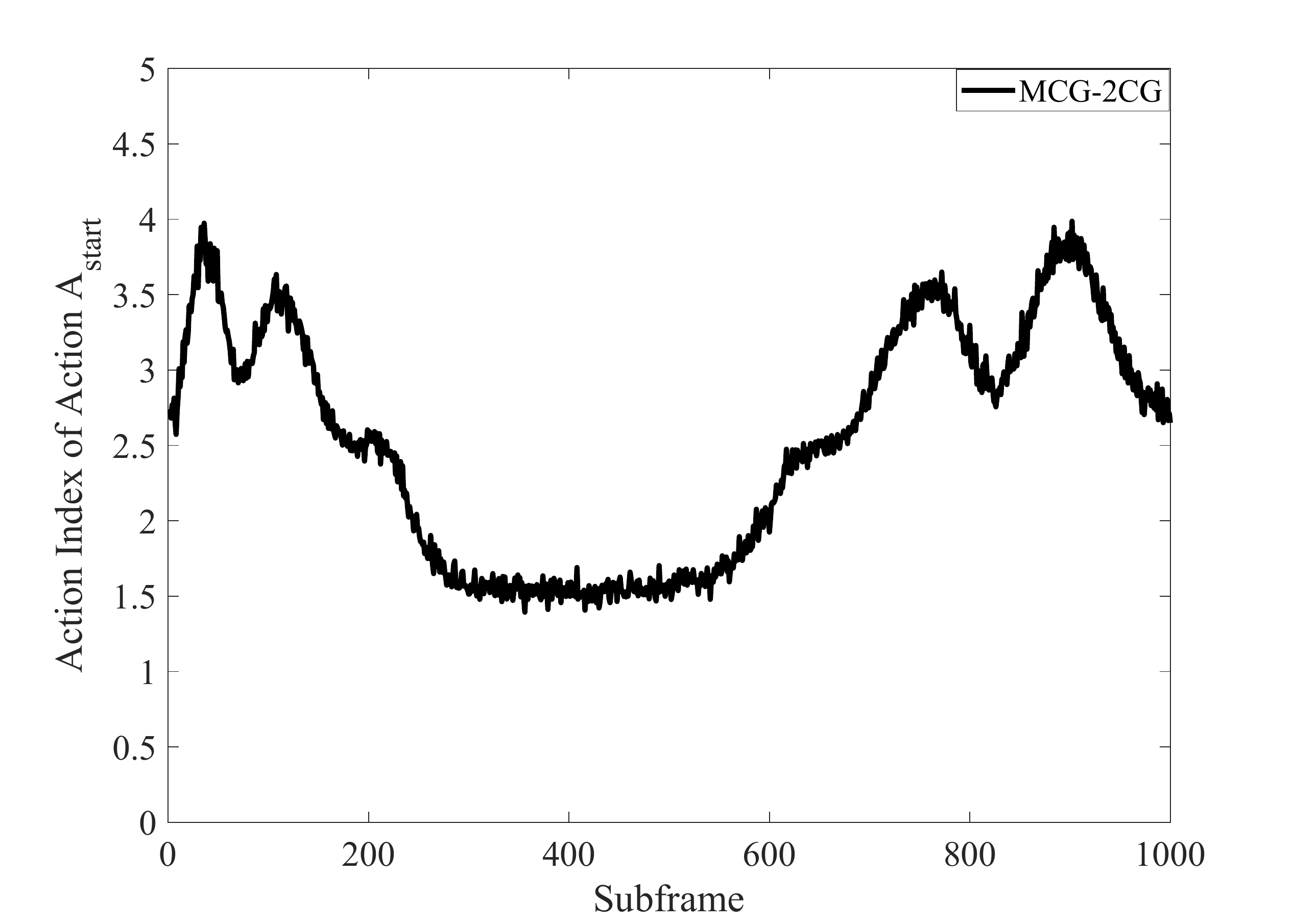}
								\vspace*{-0.4cm}
								\caption{Action index of the action $A_{\rm start}$ in the action set ${\cal A}_{\rm start}$ for MCG-GF-NOMA with $N_{\rm CG}=2.$}
								\label{fig:17}
								%   \end{minipage}
						\end{center}
					\end{figure}

					Fig.~\ref{fig:16} and Fig.~\ref{fig:17} plot the
					action index of the action $A_{\rm CTU}$ in the action set ${\cal A}_{\rm CTU}$ and the action $A_{\rm start}$ in the action set ${\cal A}_{\rm start}$ for MCG-GF-NOMA systems in heavy traffic scenario with $N_{\rm CG}=2$, respectively.
					According to the \textbf{Algorithm 1}, we could obtain the action set ${\cal A}_{\rm CTU}=\{[8, 56], [16, 48], [24, 40], [32, 32], [40, 24], [48, 16], [56, 8]\}
					$ as well as the action set ${\cal A}_{\rm start}=$ $\{[0, 1], [0, 2], [0, 3], [0, 4], [1, 2], [1, 3], [1, 4], [2, 3], [2, 4], [3, 4]\}$, which are sorted by the element in the matrix.
					In Fig.~\ref{fig:16}, we observe that the agent learns to adopt the action with a smaller number of CTUs for CG 1 and a larger number of CTUs for CG 2 around the peak traffic, e.g., $A_{\rm CTU}=[8,56]$.
					This is because the agent in the MCG-GF-NOMA scheme learns to sacrifice the successful transmission in CG 1 to alleviate the traffic congestion in CG 2 for heavy traffic regions to obtain a long-term reward.
					We also observe that the agent learns to adopt the action with the same number of CTUs for CG 1 and CG 2 around the low traffic, e.g., $A_{\rm CTU}=[32,32]$.
					This is because in a low traffic region with less traffic congestion the agent in the MCG-GF-NOMA scheme learns to guarantee the successful transmission in both the CG 1 and CG 2.
					Similarly, in Fig.~\ref{fig:17}, the agent learns to adopt the action with an earlier stating slot for CG 2 around the peak traffic, e.g., $A_{\rm start}=[0,1]$. This can guarantee the larger repetition value in CG2  to get high reliability.

					\section{Conclusion}
					In this paper, we proposed a novel MCG-GF-NOMA learning framework for attaining the long-term successfully served UEs under the latency constraint in mURLLC service, where bursty traffic of UEs was considered.
					We first designed and modeled the  MCG-GF-NOMA  system,  where we characterize each  CG using the parameters including the number of CTUs,  the starting slot of each CG within a subframe, and the number of repetitions of each CG. 
					We then characterized and analyzed the latency and reliability performances for each CG.
					We formulated the  MCG-GF-NOMA  resources configuration problem taking into account three constraints: 1) the CTU resource constraint is set to compare the MCG-GF-NOMA system with the SCG-GF-NOMA scheme; 2) the latency constraint is set to satisfy the latency requirement; and 3) the starting slot constraint is set to support various UL packet arrival times.
					Finally,  we proposed a  CMA-DDQN algorithm to balance the allocations of resources among MCGs so as to maximize the number of successful transmissions under the latency constraint,  which breaks  down the  selection  of  high-dimensional  parameters  into  multiple parallel  sub-tasks  with  a  number  of  DDQN  agents  cooperatively  being  trained  to  produce each parameter.
					% In this framework,  we  practically  simulate  the  random traffics, the   resource   configuration, the collision detection, and the data decoding procedures.
					% We use this generated simulation environment to train the DRL agents.
					\textcolor{red}{Our results have shown that the MCG-GF-NOMA framework can improve the low latency and high realibity performances in a massive URLLC scenario.
						In detail, the number of successfully served UEs in the MCG-GF-NOMA system is circa four times more than that in the SCG-GF-NOMA system, and the latency of successfully served UEs in the MCG-GF-NOMA system is circa half of that in the SCG-GF-NOMA system in high traffic scenario.
						\textcolor{red}{Our work will help to support the 3GPP evolution in terms of  1) the establishment of the theoretical foundation of MCG transmission procedure; and 2) PHY and MAC parameters configuration setup, evaluation, and optimization. Our proposed learning framework defined the observations, actions, and rewards to maximize long-term successfully served UEsunder the latency constrain, which can be standardized as the collected parameters from the environment.}
						From the perspective of performance improvement, determining the retransmission or not can be optimized in the future by considering both the different latency constraints and the future traffic congestion.
						Furthermore, a promising future direction is to cooperatively optimize networks along with the UEs’ key performance indicators (KPIs), such as power consumption and transmission delay. Such multi-objective optimization is quite challenging and should be addressed in the future.}

					\ifCLASSOPTIONcaptionsoff
					\newpage
					\fi	
					
					\bibliographystyle{IEEEtran}
					\bibliography{IEEEabrv,multipleCG}

				\end{document}